\documentstyle[epsf,12pt]{article}
\newcommand{\al}{\alpha'}
\newcommand{\de}{\partial}
\newcommand{\be}{\begin{equation}}
\newcommand{\ba}{\begin{eqnarray}}
\newcommand{\ea}{\end{eqnarray}}
\newcommand{\ee}{\end{equation}}

\newcommand{\we}{\wedge}
\newcommand{\ca}{\mathcal}
\newcommand{\lr}{\leftrightarrow}
\newcommand{\f}{\frac}
\newcommand{\s}{\sqrt}

\newcommand{\p}{\psi}
\newcommand{\bp}{\tilde{\psi}}

\newcommand{\ap}{\alpha}

\newcommand{\tap}{\tilde{\alpha}}
\newcommand{\bap}{\bar{\alpha}}
\newcommand{\tbap}{\tilde{\bar{\alpha}}}
\newcommand{\tbe}{\tilde{\beta}}
\newcommand{\bbe}{\bar{\beta}}
\newcommand{\tbbe}{\tilde{\bar{\beta}}}
\newcommand{\et}{\eta}
\newcommand{\bet}{\bar{\eta}}
\newcommand{\tet}{\tilde{\eta}}
\newcommand{\tbet}{\tilde{\bar{\eta}}}
\newcommand{\bxi}{\bar{\xi}}
\newcommand{\txi}{\tilde{\xi}}
\newcommand{\tbxi}{\tilde{\bar{\xi}}}
\newcommand{\mb}{\mathbf}
\newcommand{\ddd}{\cdot\cdot\cdot}
\newcommand{\no}{\nonumber \\}
\newcommand{\fkN}{\frac{k}{N}}
\newcommand{\la}{\langle}
\newcommand{\lb}{\rangle}
\newcommand{\ep}{\epsilon}
\newcommand{\ddbp}{\mbox{D}p-\overline{\mbox{D}p}}
\newcommand{\ddbt}{\mbox{D}2-\overline{\mbox{D}2}}
\newcommand{\ddbf}{\mbox{D}4-\overline{\mbox{D}4}}
\newcommand{\ddbz}{\mbox{D}0-\overline{\mbox{D}0}}
\newcommand{\ov}{\overline}

\begin{document}
\begin{titlepage}
\thispagestyle{empty}
\begin{flushright}
hep-th/0103021 \\
UT-927 \\
March, 2001 \\
\end{flushright}

\bigskip
\bigskip

\begin{center}
\noindent{\Large \textbf{Holomorphic Tachyons and Fractional D-branes}}\\
\bigskip
\bigskip
\noindent{
          Tadashi Takayanagi\footnote{
                 E-mail: takayana@hep-th.phys.s.u-tokyo.ac.jp} }\\

\bigskip
{Department of Physics, Faculty of Science \\ University of Tokyo \\
\medskip
Tokyo 113-0033, Japan}

\end{center}
\begin{abstract}
We study tachyon condensation on brane-antibrane systems in 
orbifold theories
from the viewpoint of boundary string field theory. We show that the
condensation of holomorphic tachyon fields generates various 
fractional D-branes. The boundary N=2 supersymmetry in the world-sheet 
theory
ensures this result exactly. Furthermore, our results are consistent with 
the twisted RR-charges from detailed calculations of boundary states.
We also discuss the generation of RR-charges due to holomorphic 
tachyon fields on multiple brane-antibrane pairs in flat space.
\end{abstract}
\end{titlepage}

\newpage

\section{Introduction}
\setcounter{equation}{0}
\hspace{5mm}
Tachyon fields naturally appear in open string theory if we consider 
various 
configurations of D-branes. For example, brane-antibrane systems 
\cite{BA1,BA2,Sen985} 
 and non-BPS D-branes \cite{BeGa} in Type II superstring theory indeed 
 have tachyon fields. Since the presence of tachyon means the instability 
 of the
 system, the condensation of tachyon is very important 
 to know the dynamical aspects of string theory.
 
Recently, tachyon condensation in open string theory has 
been intensively studied, pioneered by Sen (for a review see \cite{Sen994})
\footnote{For earlier work on tachyon condensation see \cite{Hal}.}. 
Sen conjectured that if the tachyon fields on these unstable D-brane 
systems condense into the bottom of the tachyon potential, then the 
negative energy density exactly cancels the D-brane tension \cite{Sen985}. 
After this conjecture was proposed, 
the tachyon potentials for various unstable brane systems have been 
studied by applying open string field theories and the off-shell structures 
for various unstable brane systems have been revealed.
Calculations with a good approximation called level truncation scheme
have been performed in Witten's cubic 
string field theory \cite{Witten} and Berkovits's superstring field theory 
\cite{Be}. The obtained tachyon potentials agree with 
the Sen's conjecture (for example see \cite{Cubic} and also refer to 
\cite{Oh} for a review of 
string field theory approach to tachyon condensation).
 
Another open string field theory which has been applied to tachyon 
condensation is the boundary string field theory (or background independent 
open string field theory \cite{Witten2,Sh}). In this theory one has 
only to discuss a finite number of string fields because the string fields 
which have expectation values are considered 
to correspond to only the relevant and marginal perturbations 
on the world-sheet. 
For example, one can 
calculate the exact tachyon potential \cite{GeSh,KuMaMo}.
Though the validity of 
this truncation has not been 
proved completely, one can compute tachyon condensation for specific 
tachyon fields without any approximation and the result agrees with the 
Sen's conjecture exactly \cite{KuMaMo}. Motivated by the previous results 
on the world-sheet $\sigma$-model approach \cite{Ts}, 
this formulation has 
also been
generalized for non-BPS 
D-branes \cite{KuMaMo2} and brane-antibrane systems 
\cite{Ho,KrLa,TaTeUe} in superstring theory.

Most of these recent developments in open string field theories
 are restricted to unstable D-brane 
systems in non-compact flat backgrounds. However, if one wants to know the 
geometrical 
aspects of tachyon condensation, one should challenge curved backgrounds. 

Some results of tachyon condensations in curved space 
have already been obtained. For example, tachyon condensation as marginal 
deformations \cite{Sen988,FrGaLeSt,matsuo} has been studied for 
${\bf Z}_2$-orbifolds
 \cite{Sen9812,MaSe,NaTaUe}.  
The condensation of holomorphic tachyon fields has been discussed in more 
general (Ricci flat) K\"ahler manifolds \cite{OzPaWa,Tatar,Ho}. For tachyon 
condensation in SU(2) WZW model see also 
\cite{HiNoTa,HiNoSu}. The approach which utilizes noncommutative geometry 
\cite{NCS} also has been applied to various compact spaces 
\cite{HiNoTa,BaKaMaTa,MaMo}.

Investigations of this problem may also be useful for the 
understanding of substringy geometry for D-branes, which is called 
``D-geometry" (for example see reviews \cite{Do} and 
references therein). Since the condensation of topologically non-trivial 
tachyon field generates lower dimensional D-branes \cite{Sen988,Witten3}, 
one can regard a D-brane roughly as a tachyon field on higher 
dimensional space.
If one handles the tachyon fields in boundary string field theory (BSFT),
 this will 
give another stringy description of D-branes and this will be useful to 
obtain 
D-geometry.

If we would like to get a BPS D-brane from a brane-antibrane system,
 it is natural to require that the tachyon field $T$ should be 
 holomorphic \cite{OzPaWa,Tatar,Ho}. 
 This fact seems to be correct in BSFT if one takes the large volume 
 limit because 
 the world-sheet theory becomes localized at $T=0$ \cite{HaKuMa} after 
 the tachyon condensation. Then this equation gives the 
 ``holomorphic cycle'' 
 (or divisor) on 
 which the BPS 
 D-brane \cite{BeBeSt} is wrapped \cite{Ho}. If one would like to discuss 
 this requirement in the world-sheet theory, holomorphy of tachyon 
 fields 
 is equally stated as the boundary (B-type) ${\ca{N}}=2$ supersymmetry 
 \cite{OoOzYi} on the world-sheet \cite{Ho}. 
 On the other hand, if we 
 consider
 the backgrounds where stringy corrections do exist, then the above 
 arguments
  will be modified. Therefore the investigation of tachyon 
  condensation
 in BSFT with the ${\ca{N}}=2$ supersymmetry is very 
 interesting in the stringy regions.

As a first step of this, in this paper we discuss tachyon condensations in 
orbifold theories from the viewpoint of boundary string field theory. 
We consider tachyon 
fields which preserve the boundary B-type ${\ca{N}}=2$ supersymmetry 
(``holomorphic tachyon"). After the
 tachyon condensation we obtain various fractional D-branes. We can identify
  the decay products completely by combining the boundary string field 
  theory 
with some results from boundary state calculations. From this argument we
 obtain intriguing identities for the characters of the discrete groups which
  define orbifolds. The world-sheet extended
 supersymmetry ensures these results of tachyon condensation exactly. 
Further if we resolve the 
orbifold singularities, then the final states are regarded as BPS D-branes 
which are wrapped on 
various holomorphic cycles. Thus we see again the correspondence between BPS 
D-branes and holomorphic tachyon fields. In this paper 
 we mainly discuss only ${\bf Z}_N$-orbifolds for simplicity.

The paper is organized as follows. In section 2, we first review some known 
facts on the relation between tachyon condensation in brane-antibrane systems 
and ${\ca{N}}=2$ supersymmetry. Further we discuss tachyon 
condensation for multiple brane-antibrane pairs and generation of RR-charges
on flat space.
 In section 3, we investigate tachyon condensation on orbifolds from the 
 viewpoint of boundary string field theory. In section 4, we conclude with 
 some future directions. In appendix A, we show the explicit 
 calculations of boundary
 state. 

\section{Tachyon Condensation in BSFT and Holomorphy of Tachyon}
\setcounter{equation}{0}
\hspace{5mm}
In this section we first review tachyon condensation on brane-antibrane
 systems in flat background within the framework of BSFT 
 \cite{Witten2,Sh,GeSh,KuMaMo,KuMaMo2,Ho,KrLa,TaTeUe} and we next 
 investigate
 the generation of D-brane charges from various brane-antibrane systems. 
 We obtain
 the topological configurations which generalize the Atiyah-Bott-Shapiro 
 construction \cite{ABS}.
 In particular we are interested in the relation between the boundary 
 ${\ca{N}}=2$ supersymmetry and tachyon condensations, which 
was first discussed in \cite{Ho}. Through the paper we use the language
 of Type IIA theory, but all the arguments can be applied
 to Type IIB theory straightforwardly.

\subsection{Tachyon Condensation on a Brane and an Antibrane}
\hspace{5mm}
In BSFT for superstring, ${\ca{N}}=1$ superconformal symmetry is preserved
 in the bulk of world-sheet, but its conformal symmetry
 is broken at the boundary due to boundary 
 interactions. In other words, only the boundary can be off-shell and the 
 open string fields are expressed as boundary interactions. Then the 
 boundary 
 interactions which describe the tachyon condensation should 
 naturally preserve
 ${\ca{N}}=1$ 
supersymmetry. To realize this 
supersymmetry 
one 
needs extra fermionic fields \cite{Witten3,HaKuMa}
on the boundary and this freedom corresponds to Chan-Paton factors of 
non-BPS D-branes and brane-antibrane systems. They are called boundary 
fermions and we write them by $\eta,\bar{\eta}$ (complex fermion) 
for a brane-antibrane and $\eta$ (real fermion) for a non-BPS D-brane.
Then the world-sheet 
action $I$ for a 
brane-antibrane 
system in flat space is given by \cite{HaKuMa,Ho,KrLa,TaTeUe}
\begin{eqnarray}
I &=& I_0+I_B,\\
I_0 &=& \frac{1}{2\pi\al}\int_{\Sigma}d^2 w[\partial_w X^{\mu}
\partial_{\bar{w}}X_{\mu}+\psi_{L}^{\mu}\partial_{\bar{w}}\psi_{L\mu}
+\psi_{R}^{\mu}\partial_{w}\psi_{R\mu}],\\
\label{eq10}
I_B &=& \int_{\partial\Sigma}d\tau d\theta\left[-\bar{{\bf \Gamma}}
D_{\theta}
{\bf \Gamma}
+\frac{1}{\sqrt{2\pi}}{\bf \Gamma}T({\bf X}^a)+\frac{1}{\sqrt{2\pi}}
\ov{T({\bf X}^a)}{\bf \bar{\Gamma}}\right], \label{bi}
\end{eqnarray}
where $w,\bar{w}$ denote the coordinates of (Euclidean) world-sheet 
$\Sigma$ and 
$\tau=w+\bar{w}$ denotes the boundary coordinate; we define $X^\mu
=X_{L}^\mu+X_{R}^\mu,\  
\psi_{R}^{\mu},\ \psi_{L}^{\mu}\ (\mu=0\sim 9)$ as the familiar 
bosonic and fermionic (left-moving and right-moving) fields on 
the world-sheet. The tachyon field 
$T({X}^a),\ov{T({X}^a)}$ depends only on the coordinates ${X}^a$ which are
along the world-volume of the brane-antibrane.

Here we have used ${\ca N}=1$ superspace formulation at the boundary
 of world-sheet as follows
\begin{eqnarray}
\left\{\begin{array}{lcl}
{\bf X}^{\mu} &=& X^{\mu}+2i\theta\psi^{\mu}\ \ \ \ (\psi^{\mu}
\equiv\f{1}{2}(\psi_{R}^{\mu}+\psi_{L}^{\mu})|_{\de\Sigma}),\\
{\bf \Gamma} &=& \eta+\theta F,\\
\bar{{\bf \Gamma}} &=& \bar{\eta}+\theta\bar{F},\\
D_{\theta} &=& \frac{\partial}{\partial\theta}
+\theta\frac{\partial}{\partial\tau}.\end{array}\right.
\end{eqnarray}
Note that ${\bf \Gamma}$ is the ${\ca N}=1$ superfield for 
the boundary fermion $\eta$.

If we write the boundary interactions $I_B$ in the component form and 
integrate out the auxiliary fields $F,\bar{F}$, then we get: 
\begin{eqnarray}
\label{eq6}
I_B &=& \int_{\de\Sigma} d\tau \left[\bar{\eta}\dot{\eta}-i
\sqrt{\frac{2}{\pi}}
\bar{\eta}\psi^{\mu}\de_{\mu}\bar{T}
+i\sqrt{\frac{2}{\pi}}\psi^{\mu}\eta {\de_{\mu}T}
+\frac{1}{2\pi}\bar{T}T\right].
\end{eqnarray}

{}From this it is easy to see that after the quantization of boundary 
fermions
\ba
\{\eta,\bar{\eta}\}=1,
\ea 
the Chan-Paton factors $\sigma_+=\f{1}{2}(\sigma_1+i\sigma_2),\ 
\sigma_-=\f{1}{2}(\sigma_1-i\sigma_2)$ and $\sigma_3$ correspond to 
$\bar{\eta},\eta$ and $[\bar{\eta},\eta]$, respectively. This explains 
the correct degree of freedom of Chan-Paton factors ($2\times 2$ matrices)
 for a brane-antibrane.
The above action includes only the perturbations which represent the 
tachyon field $T,\bar{T}$. Furthermore, one can also incorporate the gauge 
fields which correspond to the Chan-Paton factors $1$ and $\sigma_3$, but 
we
 will set these fields to zero in this paper. 

Also the world-sheet action for non-BPS D-branes can be easily obtained 
if one applies the descent relation \cite{Sen9812}. This relation says
 that if one performs the ${\mb{Z}}_{2}$ projection of the boundary
  interactions ${\bf \Gamma}=\bar{{\bf \Gamma}}$ for a brane-antibrane,
   then one gets those for a non-BPS D-brane \cite{HaKuMa,KuMaMo2}.

Now let us require ${\ca{N}}=2$ world-sheet supersymmetry. For example, 
 this supersymmetry is preserved for Calabi-Yau compactifications 
  as is well-known. In order to investigate D-branes in these examples,
 it is natural to consider a boundary analog of such an extended
 supersymmetry, though this is not generic. If we get to an 
 on-shell
 point after the tachyon condensation, then this supersymmetry will be 
 enhanced into ${\ca{N}}=2$ boundary superconformal symmetries, which are 
classified into
 A-type and B-type superconformal symmetry \cite{OoOzYi}. 
 These coincide with the
 classification of BPS D-branes in Calabi-Yau spaces in the large volume 
 limit \cite{BeBeSt}. Therefore it will be particularly interesting
  to consider the boundary interactions which preserve this symmetry.
  Then what kinds of tachyon fields 
 satisfy this requirement? It was pointed out in the 
 paper
  \cite{Ho} that the B-type supersymmetry is not broken if one considers 
 {\it holomorphic tachyon field }for brane-antibrane systems.
 
 More concretely, the boundary interaction
   which preserves B-type ${\ca{N}}=2$ supersymmetry (below we will omit 
   the word `B-type' and simply call this ${\ca{N}}=2$ supersymmetry) can 
be written \cite{Ho} as follows 
(for earlier relevant work see also \cite{Wa})
\ba
I_B &=& -\int_{\de\Sigma} d\tau d\theta d\bar{\theta}
{\bf{\Gamma}}\bar{{\bf \Gamma}}+\int_{\de\Sigma} d\tau d\theta \frac{1}
{\sqrt{2\pi}}{\bf \Gamma}T({\bf Z}^i)+(\mbox{h.c.}), \label{n2b}
\ea
where we have employed ${\ca{N}}=2$ boundary superspace 
$(\tau,\theta,\bar{\theta})$ and the boundary fermionic 
chiral and antichiral superfield 
${\bf \Gamma},\ov{{\bf \Gamma}}$ are defined in our conventions as
\ba
{\bf \Gamma}&=&-\f{i}{\s{2}}\eta+\theta F-\f{i}{\s{2}}\theta\bar{\theta}
\de_{\tau}\eta, \no
\ov{{\bf \Gamma}}&=&\f{i}{\s{2}}\bar{\eta}+\bar{\theta} \bar{F}
-\f{i}{\s{2}}
\theta\bar{\theta}
\de_{\tau}\bar{\eta}.
\ea

Note that the tachyon field $T(Z^i)$ 
depends only on the holomorphic coordinates $Z^i=X^{2i-1}+iX^{2i}$ 
along the world-volume in order to preserve ${\ca{N}}=2$ boundary 
supersymmetry.

The most interesting issue of ${\ca{N}}=2$ supersymmetry is the fact that 
the boundary 
superpotential term $\sim \int_{\de\Sigma} 
d\tau d\theta {\bf \Gamma}T({\bf Z}^i)$ is not renormalized 
as argued in \cite{Ho}. On the other hand the kinetic term for the boundary 
fermion is included in the boundary D-term and will receive quantum 
corrections. We assume that the contributions from the D-term are not 
singular
 and therefore the potential term dominates the D-term after the 
 tachyon condensation $|T_{i}|\to \infty$.
 For example, let us assume the following 
holomorphic tachyon field on a $\ddbt$ which is extended 
in $Z^1(\equiv Z)$ 
direction \cite{Ho}:
\ba
T(Z)=\sum_{k=0}^{p}a_{k}\cdot Z^k=a_{p}\prod_{k=1}^{p}(Z-z_{k}) \label{tz}.
\ea
Then the values of $\{z_{k}\}$ are not renormalized. As Sen and Witten 
argued in \cite{Sen988,Witten3}, if the tachyon field which has a 
topological charge 
does condense, then the corresponding lower dimensional D-branes 
are generated. 
In our example the tachyon field (\ref{tz}) has the winding number $p$ 
and thus $p$ D0-branes should be produced at each point $z_{k}$. 

Let us see this in BSFT. In superstring theory the spacetime action $S$ of 
BSFT is argued to be identified with the disk partition function $Z_{disk}$ 
\cite{Ts,KuMaMo2,Ho,KrLa,TaTeUe}
\ba
S=Z_{disk}=\int [DX] [D\psi] [D\eta][D\bar{\eta}]\exp({-I_{0}-I_{B}}). 
\label{bsft}
\ea
As the tachyon condenses infinitely $a_{p}\to \infty$, the path integrals 
around 
the $p$ fixed points $Z=z_{k}$ give dominant contributions to $Z_{disk}$. 
Then
 the partition function becomes $p$ times that for $p=1$ case \cite{Ho}. 
 On the other hand, the boundary perturbation for $p=1$ can be treated within
  a free theory. Using the results in \cite{KuMaMo2,KrLa,TaTeUe,Ho}, one can 
show 
\ba
\f{Z(a_{1}=0)\times (\mbox{Vol})^{-1}}{Z(a_{1}=\infty)}=\f{1}{2\pi^2\al}
=\f{T_{D2-\bar{D2}}}{T_{D0}},
\ea
where $T_{D0}$ and $T_{D2-\bar{D2}}$ denote the tension of a D0-brane and of 
a
$\ddbt$, respectively; $\mbox{Vol}$ denotes the volume of the D2-brane
 world-volume. Thus after the condensation of tachyon field (\ref{tz}),
  $p$ D0-branes are produced as expected.

Another way to see this is to compute the RR-couplings of a $\ddbt$.
 As discussed in \cite{KeWi,KrLa,TaTeUe,AlItOz}, those couplings for 
a $\ddbp$ system 
in BSFT are
written by using Quillen's superconnection \cite{Qu} if we ignore the
the contributions from non-abelian transverse scalars \cite{TaTeUe,Te}.
They are given by the following formula
\ba
S&=& T_{Dp}\ \mbox{Str}\ \int C\wedge \exp (2\pi\al\cal{F}) \no
2\pi\al\cal{F}&=&\left(
	\begin{array}{cc}
	2\pi\al F^{(1)}-\bar{T}T & (i)^{\f32}\s{2\pi\al}\ \overline{DT} \\
	(i)^{\f32}\s{2\pi\al}\ DT & 2\pi\al F^{(2)}-T\bar{T}
	\end{array}
\right), \label{wz}
\ea
where $\mbox{Str}$ is supertrace and $\cal{F}$ is the field strength
 of superconnection \footnote{We have replaced $T$ with $\bar{T}$ in the
  reference \cite{TaTeUe}. Note also that a factor $-1$ in front of
  $DT$ is different from eq.(4.8) in \cite{TaTeUe}.
   This is because here we assume $T$ anticommutes with any odd-forms.}.
Then let us compute the RR-coupling which represents the D0-brane
charge in the previous example. As shown in \cite{Qu}, continuous
deformations of the tachyon field do not change the result.
 Therefore we can restrict the form of tachyon field to
\ba
T(Z)=a_p\cdot Z^p.
\ea
Then the integration in (\ref{wz}) along the coordinate $Z^1$ does not
 depend on $a_p$ and we obtain the following RR-coupling 
\ba
S_{RR}&=&(i2\pi\al)T_{D2}\int C_{D0}\we dT\we \overline{dT}\ e^{-T\bar{T}}\no
&=&4\pi\al p^2 T_{D2} \int C_{D0}\int_{0}^{\infty}2\pi r dr\ r^{2p-2}\ 
e^{-r^{2p}}\no
&=&p\ T_{D0}\int C_{D0},
\ea
where we have used the relation $T_{D0}=(2\pi)^p(\al)^{\f{p}{2}}T_{Dp}$; 
the 1-form
 field $C_{D0}$ denotes the RR-field which couples to D0-branes. Thus we
  get $p$ units of D0-brane RR-charge matching with the above result.

Next we would like to comment on the higher dimensional generalization. If
the tachyon field $T$ depends only on one coordinate (for example, $Z^1$),
 then the generalization is trivial. More generally,
let us consider the tachyon field $T(Z^1,Z^2,\ddd,Z^n)$ on a 
 $\mbox{D}2n-\overline{\mbox{D}2n}$. If the holomorphic function $T$ is
 reducible as $T=T^{(1)}\cdot T^{(2)}\ddd T^{(q)}$, then we obtain the sum 
 of the decay products each corresponding to 
 $T=T^{(1)},T=T^{(2)},\ddd,T=T^{(q)}$ \cite{Ho}. 
 Therefore we can 
 assume the function $T(Z^1,Z^2,\ddd,Z^n)$ is irreducible. 
 Then we will obtain a 
 D$(2n-2)$-brane wrapping on a codimension two hypersurface $T=0$. However,
  this may be problematic.
 In general this configuration of the `curved' D-brane
 seems to be unstable in spite of its holomorphy
 since the D-brane is put in flat space and cannot wrap any cycles. It would
  be interesting to investigate this further, though we mainly discuss the
  generation of D0-brane charges in this paper.

Before closing this subsection, let us ask what will happen 
if we do not assume the boundary ${\ca{N}}=2$ 
supersymmetry. First, one can produce lower dimensional non-BPS D-branes. 
This requires a kink-like tachyon field and is not holomorphic. 
Second, one will also be able to produce D0-branes and anti D0-branes 
at the same time. 
For example, let us consider the following tachyon field on a $\ddbt$
\ba
T(Z,\bar{Z})=a_{q+p,q}Z^{q+p}\bar{Z}^{q}. \label{tba}
\ea
 In the same way as the above RR-charge computation, one can calculate 
 D0-brane charge of this configuration. The result is $p$ times 
 that of a D0-brane, which can be also seen intuitively from the fact 
 that the tachyon 
 field (\ref{tba}) has the winding number $p$. One may hastily conclude that 
 the configuration (\ref{tba}) generates a system of $(q+p)$ D0-branes and 
 $q$ anti D0-branes after the tachyon condensation. In fact this boundary
  interaction (\ref{tba}) breaks 
 ${\ca{N}}=2$ supersymmetry and thus should be renormalized. Therefore we 
 can argue that a system of $(q+p)$ D0-branes and $q$ anti D0-branes for 
 any 
 $q$ will be produced in a certain limit of the following configurations
\ba
T(Z,\bar{Z})=\sum_{q=0}^{\infty}a_{q+p,q}Z^{q+p}\bar{Z}^{q}.
\ea
Generally, these are highly interactive theories and it will be difficult to 
analyze further.

\subsection{Tachyon Condensation on Multiple Branes and Antibranes}
\hspace{5mm}
The above formulation can be generalized for multiple 
brane-antibrane systems. The path-ordered formulation for these 
was given in \cite{TaTeUe}. If one wants to construct the corresponding
 ${\ca{N}}=1$ boundary interaction, one has only to include more
 than one boundary fermions \cite{Ho,KrLa}. We write the superfields
 for them as ${\bf \Gamma}_i,\ov{{\bf \Gamma}_i}\ \ (i=1,2,\ddd,n)$. 
 The quantization of boundary
 fermions $\eta_{i}$ is written by
\ba
\{\eta_i,\bar{\eta_{\bar{j}}}\}=\delta_{i\bar{j}}.
\ea
Comparing this with the algebra of $\gamma$ matrices
 $\gamma_{1},\ddd,\gamma_{2n}$:
\ba
\{\gamma_{a},\gamma_{b}\}=2\delta_{ab},
\ea
we get the correspondence
\ba
\eta_i \lr \gamma^{+}_{i}\equiv\f{1}{2}(\gamma_{2i-1}+i\gamma_{2i}),\ \ \ 
\bar{\eta}_{\bar{i}} \lr \gamma^{-}_{\bar{i}}\equiv\f{1}{2}
(\gamma_{2i-1}-i\gamma_{2i}).
\ea
Then we can get $2^n \times 2^n$ Chan-Paton matrices which
 corresponds to $2^{n-1}$ branes and $2^{n-1}$ antibranes
  even though for the general number of branes and antibranes,
   ${\ca{N}}=1$ boundary superspace formulation is not known.
    In this formalism, the boundary interactions are expanded as 
\ba
I_{B}=
\int_{\partial\Sigma}d\tau d\theta\left[-\bar{{\bf \Gamma}_i}D_{\theta}
{\bf \Gamma}_i
+\f{{\bf \Gamma}_{i}T_{i}({\bf X}^a)}{\sqrt{2\pi}}+
T_{ijk}({\bf X}^a){\bf \Gamma}_i{\bf \Gamma}_j
{\bf \Gamma}_k+
T_{\bar{i}jk}({\bf X}^a)\bar{{\bf \Gamma}}_{\bar{i}}{\bf \Gamma}_j
{\bf \Gamma}_k+
\cdot\cdot+\mbox{(h.c.)}\right] , \nonumber 
\\
\ea
where we have omitted the summation over the indices $i,j,k,\bar{i}$.
Note that in the above equation the fields $T_{\bar{i}},
T_{ijk},T_{\bar{i}jk},\ddd$ represent
 non-abelian tachyon fields on $2^{n-1}$ brane-antibrane pairs. Here the
  non-abelian gauge fields are neglected again and these correspond to the
   boundary interactions which include even number of boundary
    fermionic superfields. 

If we are interested in ${\ca{N}}=2$ boundary supersymmetry, then
 the above boundary interactions should be constrained. The boundary
 interactions which represent tachyon fields should be in the boundary
superpotential terms $\sim \int_{\partial\Sigma}d\tau d\theta\ W$. 
Therefore (i) tachyon fields should be {\it holomorphic} (or depend
 only on $Z_i$) and (ii) the potential terms should involve 
 {\it no anti-chiral superfields 
$\bar{\bf \Gamma}_{\bar{i}}$}. For example, the second requirement does not
 allow the field $T_{\bar{i}jk}$.

Now let us consider tachyon condensation on $2^{n-1}$ D($2n$)-branes and 
$2^{n-1}$ anti D($2n$)-branes. In such a system there should be decay modes 
which 
generate BPS D0-branes following the general arguments in K-theory 
\cite{Witten3}. We assume the following ${\ca{N}}=2$ boundary
 interaction \cite{Ho} for simplicity
\ba
I_B &=& -\int_{\de\Sigma} d\tau d\theta d\bar{\theta}\sum_{i}
{\bf{\Gamma}}_i\bar{{\bf \Gamma}_i}+\int_{\de\Sigma} d\tau d\theta \frac{1}
{\sqrt{2\pi}}\sum_{i}{\bf \Gamma}_iT_i({\bf Z})+(\mbox{h.c.}), \label{bim}
\ea
where the tachyon fields $T_i(Z)$ depend only on the holomorphic
 coordinate  ${ Z}^1,\ddd,{ Z}^n$
 of the world-volume
. Note that if $n=1$ or $2$, then the general 
${\ca{N}}=2$ boundary interaction can be written as the above form. As 
argued in \cite{Ho} the condensation of these tachyon fields generally
 produces a D-brane wrapped on the intersection of hyper-surfaces
  $T_i({\bf Z})=0$. Below we would like to investigate this further. The 
results will be useful in the next section.

We first turn to the RR-coupling which corresponds to
 the D0-brane charge. The 
non-abelian tachyon field $T$ can be written by
\ba
\left(
	\begin{array}{cc}
	0 & \overline{T} \\
	T & 0
	\end{array}
\right)
=\sum_{i}\gamma^+_{i}T_{i}+\sum_{i}\gamma^-_{\bar{i}}\bar{T}_{\bar{i}},
\label{takm}
\ea
where note that $\gamma$ matrices here are not projected into the Weyl
 representation. Notice that we regard the tachyon fields as holomorphic
 if $T_{i}$ are holomorphic functions. Generally this does not mean that
 the non-abelian tachyon field $T$ in the above matrix is holomorphic 
 in an ordinary sense.

Putting this into eq.(\ref{wz}), we get the following RR-coupling 
 in BSFT:
\ba
S_{RR}&=&T_{D(2n)}\ \mbox{Str} \int C_{D0}\ \exp(-\sum_{i=1}^{n}|T_i|^2)
 \we \exp
[\s{-2i\pi\al}\sum_{i=1}^{n}(\gamma^{+}_i dT_i +\gamma^{-}_{\bar{i}} 
d\bar{T}_{\bar{i}})]\no
&=&T_{D(2n)}\ (-2i\pi\al)^n\times \f{1}{(2n)!}\int C_{D0}
\exp(-\sum_{i=1}^{n}|T_i|^2)
\we \mbox{Tr}[\gamma_{2n+1}(\sum_{i=1}^{n}\gamma^{+}_i dT_i +
\gamma^{-}_{\bar{i}} 
d\bar{T}_{\bar{i}})^{2n}]\no
&=& (2i\pi\al)^n\ T_{D(2n)}\int \exp(-\sum_{i=1}^{n}|T_i|^2)\ C_{D0}\we
\prod_{i=1}^{n}dT_{i}\we d\bar{T}_{\bar{i}}, \label{mwz}
\ea
where the chirality matrix 
$\gamma_{2n+1}=(i)^{-n}\gamma_1\gamma_2\ddd\gamma_{2n}$ was inserted 
in order to replace the supertrace Str with the ordinary trace Tr. Note that 
 there are no RR-charges other than D0-branes produced from the tachyon 
 fields
 (\ref{takm}) because of the trace over $\gamma$ matrices.
 
If we assume $T_i$ depends only on $Z_i$, then the above integrations are 
divided into $n$ independent parts. If one sets the degree of $T_i(Z_i)$ 
is $p_i$, then the result is given by 
\ba
S_{RR}&=&(\prod_{i=1}^{n}p_i)\cdot T_{D0}\int C_{D0}.
\ea
Thus we can conclude that $(\prod_{i=1}^{n}p_i)$ D0-branes are generated in 
this case. Furthermore one can show that this configuration has a 
correct tension in BSFT.
To see this one has only to note that the partition function $Z_{disk}$ 
is also divided into $n$ independent path integrals for each direction 
$Z_1,\ddd,Z_n$
\ba
Z_{disk}=\prod_{i=1}^{n}Z^i_{disk}.
\ea
Then using the previous result of tachyon condensation on a $\ddbt$, 
it is easy to see the resulting tension is ($\prod_{i=1}^{n}p_i$) times 
that of a D0-brane. In particular the configuration $p^1=\ddd=p^n=1$ 
generates
 a BPS D0-brane and corresponds to Atiyah-Bott-Shapiro \cite{ABS} 
construction of K-theory charges. 

Let us turn to other configurations. For simplicity, we set $n=2$ and 
consider 
a system which is consist of two $\mbox{D}4$-branes and two anti D4-branes. 
We consider the following holomorphic tachyon fields for generic examples
\ba
T_1(Z_1,Z_2)=(Z_1)^p(Z_2)^q-a,\ \ \ \ \
T_2(Z_1,Z_2)=(Z_1)^r(Z_2)^s-b, \label{t2}
\ea
where $p,q,r,s$ are non-zero integers and we assume $ps-qr\neq 0$. 

First note that the above configurations include only 0-branes if either
$a$ or $b$ is not zero. This is
 because if one calculates the disk partition function in BSFT, one always 
finds the factor $\exp(-\sum_{i=1}^{2}T_i\bar{T}_i)$ and this means that
 the degree of freedom will be localized at the points (0-branes) defined 
 by the equations $T_1=T_2=0$. One can calculate the number of the points 
 and the result\footnote{To see this, let $l$ be 
the g.c.d. of $p$ and $r$ as $p=l\cdot \ap$
 and $r=l\cdot \beta$. Then one obtains $|q\beta-s\ap|$ solutions about
  $Z_2$ as $(Z_2)^{q\beta-s\ap}=a^{\beta}\cdot b^{-\ap}$. 
After we insert this in $T_1=T_2=0$ again, we get $l$ solutions about $Z_1$. 
Thus
 we get $|ps-rq|$ solutions.}  is given by
  $|ps-rq|$ . This shows the total number of generated D0-branes and 
  anti D0-branes is $|ps-rq|$ because one fixed point 
  gives the tension of a D0-brane (or anti D0-brane). On the other hand, 
  one can also calculate 
 the D0-brane RR-charge of these configurations with an appropriate 
 change of the variables in the integration (\ref{mwz}). The result is
 $(ps-qr)$ times that of a D0-brane. Thus we can conclude that 
the above tachyon fields (\ref{t2}) generate $(ps-rq)$ BPS D0-branes unless
 $a=b=0$. Mathematically one can say that the integration (\ref{t2}) counts 
 the
 number of the (localized) solutions to the algebraic equations 
 $T_{i}=0$ and this result will hold for general $n$ and nonsingular 
 polynomials $T_{i}$.

Next we consider the singular cases $a=b=0$. The D0-brane RR-charge is again
 given by $(ps-rq)$. After the condensation of these
 tachyon fields, 2-branes will also be generated since the equations 
 $T_1=T_2=0$ are satisfied for $Z_1=0$ or $Z_2=0$. These 2-branes should be
  $\ddbt$ systems because this configuration does not have D2-brane 
  RR-charge.
The generation of $\ddbt$ is not so surprising. If one assumes $p=r,\ q=s=0$,
 then this configuration corresponds to the decay 
 into $p$ pairs of $\ddbt$ at $Z_1=0$ as can be seen easily
 by using $U(2)$ rotational 
 symmetry of
 $({\bf \Gamma}_1, {\bf \Gamma}_2)$. Though we cannot determine how many
 $\ddbt$ systems will be produced for general $(p,q,r,s)$, it will be 
 interesting to note that a system which is generically D0-branes can become
 higher dimensional branes for singular points in the field space of BSFT.

Finally we would like to comment on the relation between various tachyon 
condensation modes and ${\ca{N}}=2$ boundary (B-type) supersymmetry. 
In the above
 arguments on tachyon condensations in brane-antibrane systems, 
 we have not observed the generation of D0-branes and anti D0-branes 
 at the same time\footnote
 {Of course, if one adds more boundary fermions with preserving 
 ${\ca{N}}=2$ supersymmetry, then we can obtain $\ddbz$ systems. What we 
 would like to say here
 is that we cannot obtain $\ddbz$ systems if we start from the minimal 
 number ($=2^{n-1}$) of $\mbox{D}(2n)-\overline{\mbox{D}(2n)}$ pairs.}. 
 As can be seen from this 
 example we believe that there is an intriguing correlation in general 
 backgrounds between the BPS
 nature of final objects and the holomorphy of tachyon (or
 ${\ca{N}}=2$ supersymmetry). 
 In the next section we will see another 
 evidence of this argument in orbifold theories, which give the simplest 
examples in curved spaces.

\section{Tachyon Condensation on Orbifolds}
\setcounter{equation}{0}
\hspace{5mm}
In this section we discuss tachyon condensation in brane-antibrane systems
 on orbifolds \cite{orbifold}. Mainly we consider the four dimensional
 orbifolds ${\mb{C}}^2/{\bf Z}_N\ \ (N\geq 2)$, but the similar 
 arguments will be applied to higher dimensional examples or more 
 complicated orbifold projections. The relation between the tachyon 
 condensation in these systems and the equivariant K-theory was discussed 
 in \cite{Witten3,GC}. The tachyon condensation from the viewpoint of 
 noncommutative geometry \cite{NCS}  was also discussed for orbifolds 
 \cite{MaMo}.
 Here we investigate this in the framework of BSFT and determine what will 
 be 
generated after the tachyon condensation precisely. Before we do so, 
let us first review some 
useful facts about D-branes on orbifolds \cite{DoMo,JoMy}. 

In Type II superstring theory we can consider the
 orbifold projections ${\bf \Gamma}$
   on ${\bf C^2}$ 
 which preserve half of the bulk 
 supersymmetries\footnote{Of course the discrete
  group ${\bf \Gamma}$
 is completely different from the fermionic boundary superfield 
 ${\bf \Gamma}$, though we use the same symbol below.}. This means that the 
 discrete groups ${\bf \Gamma}$ 
 of the orbifold 
projections should be subgroups of $SU(2)$ and are known to be classified 
into A,D,E series. Geometrically, the orbifolds ${\mb{C}}^2/{\bf \Gamma}$
 can be realized 
in the neighborhoods
 of the A,D,E singularities in K3 surface. These singularities are due 
 to the 
 vanishing 2-cycles in
 K3. If they are resolved by blowing up, then one gets ALE spaces
 (see for example \cite{EgGiHa}).
 However, in string theory these singularities do not imply physical
 singularities. Indeed there are B-field fluxes (=twisted NSNS-fields) 
through the 2-cycles \cite{As} and the world-sheet instantons and various 
D-branes which wrap these cycles
 do not become tensionless. Thus the theory is not singular.

Below we concentrate on the A series for simplicity, which are equivalent 
to the familiar discrete groups ${\bf Z}_N$ . The action of ${\bf Z}_N$ is
 defined as follows:
\ba
1,g,g^2,\ddd,g^{N-1}\in {\bf Z}_N,\ \ \ (g^N=1) ,\no
g:z_1\to e^{\f{2\pi i}{N}}z_1,\ \ \ z_2\to e^{-\f{2\pi i}{N}}z_2,
\ea
where $z_1,z_2$ denote the coordinates of ${\bf C^2}$. 

Now let us turn to D-branes on ${\mb{C}}^2/{\bf Z}_N$. In this paper we 
always assume that the D-branes are particle-like in the ${\bf R}^{1,5}$ 
direction. 
Then BPS D$p$-branes exist for $p=0,2,4$. In particular D4-branes are wrapped 
on 
the whole ${\mb{C}}^2/{\bf Z}_N$. The D2-branes which are parallel to the 
$z_1$-plane or $z_2$-plane are BPS objects and can be treated with the 
world-sheet ${\ca{N}}=2$ supersymmetry. 

The open string spectrum of D$p$-branes on the
 orbifold can be given by ${\bf \Gamma}$-projection on the Chan-Paton 
 degree
 of freedom \cite{DoMo,JoMy}. In other words, D$p$-branes on 
 ${\mb{C}}^2/{\bf \Gamma}$ are classified by the group theoretical 
 representations of ${\bf \Gamma}$ action on Chan-Paton factors. For
 ${\bf \Gamma}={\bf Z}_N$, there are $N$ irreducible representations and 
we denote these by $\{\rho_{\ap}\}\ (\ap=0,1,2,\ddd,(N-1))$. The 
representation $\rho_{\ap}$ is defined as the one dimensional representation
 which gives the phase rotation $\exp(\f{2\pi i\ap}{N})$. Then we
 call a D$p$-brane which corresponds to $\rho_{\ap}$ representation
 a $\ap$-type D$p$-brane. These $N$ kinds of D-branes are the most fundamental
 D-branes. For $p=0$ they are called fractional D-branes \cite{DiDoGo}, which
  are identified with the D2-branes wrapped on vanishing 2-cycles. 
  It is known
  that vanishing 2-cycles are also classified by the irreducible 
  representations 
  and a $\ap$-type D0-brane corresponds to a D2-brane wrapped on the 2-cycle 
  $[\ap]$ \cite{DiDoGo,DiGo,Ta}. Fractional D-branes are fixed at the origin 
  $z_1=z_2=0$ and cannot move from there. The tension of each of them is
  $\f{1}{N}$ times that of a bulk D0-brane, which can move freely in the 
  orbifold. 
 On the other hand, for $p=2,4$ such a D$p$-brane has the same tension as the
  ordinary D$p$-brane since the $g$-action acts on the world-volume 
  non-locally. 
These facts can also be verified by using boundary state formalism for 
orbifold 
theories (see for example \cite{OoOzYi,HuIeNuSc,Sen98n,BiCrRo,Br,DiGo,Ta,
BiCrRo2,BeDi}) 
as we will see in the appendix A.

 All the  other D-branes in the orbifold theory are regarded 
 as linear combinations of these fundamental D$p$-branes and 
 correspond to all of the reducible
 representations as $\rho=\oplus_{\ap=0}^{N-1}c_{\ap}\rho_{\ap}\ 
 (c_{\ap}\in {\bf Z})$. For example, the regular representation 
 $\rho_{reg}=\oplus_{\ap=0}^{N-1}\rho_{\ap}$ corresponds to a bulk D0-brane. 
 Open strings between a $\ap$-type D$p$-brane and a
 $\beta$-type D$p$-brane belong to the representation $\rho_{\beta}\otimes
\rho_{\ap}^{*}$, where $*$ denotes the complex conjugation\footnote{
Note that if one changes the orientation of the open strings, then they
belong to $\rho_{\ap}\otimes\rho_{\beta}^{*}$.}. Then the
 super Yang-Mills theories called quiver gauge theories are realized on the
 world-volume of BPS D-branes as shown in \cite{DoMo}.
 
Here we are interested in the tachyon field $T$ which comes from the
open string between a $\ap$-type D$p$-brane and a $\beta$-type anti 
D$p$-brane. 
These open strings belong to $\rho_{\beta}\otimes\rho_{\ap}^{*}$ with the
 opposite GSO-projection and the $g(\in{\bf Z}_N)$ action is given by
\ba
g\ :\ T(z_1,z_2)\ \to\ e^{\f{2\pi i}{N}(\ap-\beta)}\cdot
T(e^{\f{2\pi i}{N}}z_1,e^{-\f{2\pi i}{N}}z_2). \label{tp}
\ea

\subsection{Tachyon Condensation on Orbifolds in BSFT}
\hspace{5mm}
Now let us investigate the tachyon condensation on orbifolds in BSFT. Again
 we are interested in holomorphic tachyon fields, 
 which preserve ${\ca{N}}=2$ supersymmetry.
As mentioned in the previous section, the spacetime action of BSFT 
in flat space is defined as the disk partition function eq.(\ref{bsft}).
  If the world-sheet action $I_{0}+I_{B}$ is invariant under a certain
 transformation of the world-sheet fields ${\bf X^\mu}$ and ${\bf \Gamma}$, 
 we 
can twist the theory by this symmetry. In particular if we regard 
$g\in{\bf Z}_N$ as the symmetry, then we get the BSFT action for D-branes
 on orbifolds. 

\begin{flushleft}
\noindent{\bf Generation of Codimension Two D-branes}
\end{flushleft}

Let us first turn to a $\ddbt$ pair of which world-volume is defined by
 $z_2=0$. We assume that the D2-brane is $\ap$-type and the anti D2-brane is 
 $\beta$-type. Note that the branes cannot move from $z_2=0$. If we remember 
 the $g\in{\bf Z}_N$ action (\ref{tp}), the tachyon field should be projected 
 as follows:
\ba
T(z_1)=e^{\f{2\pi i}{N}(\ap-\beta)}\cdot
T(e^{\f{2\pi i}{N}}z_1).
\ea
In BSFT this can be equally stated that the boundary interaction (\ref{n2b})
 should be invariant under the following transformation
\ba
g\ :\ {\bf\Gamma}\ \to\ e^{\f{2\pi i}{N}(\ap-\beta)}{\bf\Gamma},\ \ 
{\bf Z}_1\ \to\ e^{\f{2\pi i}{N}}{\bf Z}_1.
\ea
Then the allowed tachyon field can be given by
\ba
T(z_1)=a_{q}\cdot (z_1)^{\beta-\ap+Nq},\label{t11}
\ea
where $q$ is non-negative integers for $\beta \geq \ap$ and is positive 
integers for $\ap > \beta$. The BSFT action becomes
\ba
S&=&\int_{{\mb{C}}^2/{\bf \Gamma}} [D{\bf Z}_1][D{\bf \Gamma}]
\ \exp({-I_{0}-I_{B}})
\no &=&\f{1}{N}\int_{{\mb{C}}^2}[D{\bf Z}_1][D{\bf \Gamma}]
\ \exp({-I_{0}-I_{B}}) \no
&=&\left(\f{\beta-\ap+Nq}{N}\right)\cdot T_{D0}\int dx^{0},
\ea
where we have used the fact that the disk partition function after the 
condensation of the tachyon field (\ref{t11}) is the same as that for 
$(\beta-\ap+Nq)$ D0-branes as explained in the previous section. 
The calculation of bulk RR-charge\footnote{Here ``bulk RR-charge'' means
 the RR-charge in the untwisted-sector. Notice that there are also twisted 
RR-charges which is characteristic of orbifold theories. These charges will
 be discussed later.} is also in the same way as 
in section 2 and the result is $(\beta-\ap+Nq)/N$ times that of a BPS 
bulk D0-brane\footnote{
Note that one can also consider $g$ invariant tachyon field $T=a_{q}\cdot 
(\overline{z_1})^{\ap-\beta+Nq}$. For these the different ${\ca{N}}=2$ 
supersymmetry 
is preserved and have opposite RR-charge. Thus fractional anti D-branes will
be produced from these.}. Thus we can
 conclude that $\beta-\ap+Nq$ fractional D0-branes will be generated at the 
 point $z_1=0$. Then what
 kinds of fractional branes will be generated? To answer this question 
 completely we need the knowledge of twisted RR-charges and we will 
 return to this in the next subsection. Nevertheless we can obtain
 some hints from the above arguments. First let us set $\beta=\ap$. Then the 
mode $q=0$ corresponds to the decay into the vacuum as the tachyon condenses 
$a_{0}\to\infty$. This is impossible for other cases $\beta\neq\ap$ since 
the types of the brane and the antibrane are different and they cannot 
annihilate. Note also that the tachyon field for $\beta=\ap$ have $qN$ zeros 
and they are invariant by the geometric ${\bf Z}_N$ action even if we deform 
the tachyon field (\ref{t11}) by allowed polynomials. Then it is natural
 to identify these zeros as $q$ bulk D0-branes. For example, it is obvious 
 that they can move from $z_1=0$. Furthermore, it is not difficult to see
 that the condensation of the tachyon (\ref{t11}) will
 generate both $q$ bulk D0-branes and $(\beta-\ap)$ fractional D0-branes
  if we assume $\beta \geq \ap$. On the other hand, if we assume $\beta <\ap$
 ,then we will obtain both $(q-1)$ bulk D0-branes and $N+(\beta-\ap)$ 
 fractional D0-branes.

Next we turn to tachyon condensation on a $\ddbf$ pair.
In this case we can assume the following tachyon 
field 
\ba
T(z_1,z_2)=a_{q,r}\cdot (z_1)^{\beta-\ap+Nq}\cdot(z_1z_2)^r \label{ddb41}
\ea
We can apply the RR-coupling formula (\ref{wz}) or (\ref{mwz}) to this.
Then it is easy to see that the final state after the tachyon condensation
 consists of $(\beta-\ap+Nq+r)$ D2-branes on $z_1=0$ and $r$ D2-branes on 
 $z_2=0$, each of which corresponds to a irreducible representation.

\begin{flushleft}
\noindent{\bf Generation of Codimension Four D-branes}
\end{flushleft}

Next we consider two $\ddbf$ pairs and discuss the generation of D0-branes. 
We can use the boundary interaction (\ref{bim}) with $i=1,2$. To see the 
matrix representation of tachyon fields $T_1,T_2$ explicitly, let us use the 
standard expressions of $\gamma$ matrices
\ba
\gamma_1=
\left(
	\begin{array}{cc}
	0 & \sigma_1 \\
	\sigma_1 & 0
	\end{array}
\right)
 ,\ \ 
\gamma_2=
\left(
	\begin{array}{cc}
	0 & \sigma_2 \\
	\sigma_2 & 0
	\end{array}
\right)
 ,\ \ 
\gamma_3=
\left(
	\begin{array}{cc}
	0 & \sigma_3 \\
	\sigma_3 & 0
	\end{array}
\right)
,\ \ 
\gamma_4=
\left(
	\begin{array}{cc}
	0 & -i{\bf1} \\
	i{\bf1} & 0
	\end{array}
\right),
\ea
where $\sigma_1,\sigma_2,\sigma_3$ denote Pauli matrices. Then the 
non-abelian tachyon field $T$ is given by
\ba
T=
\left(
	\begin{array}{cc}
	\ov{T_2} & T_1 \\
	\ov{T_1} & -T_2
	\end{array}
\right). \label{t4}
\ea

Now we assume that the two D4-branes and two antiD4-branes correspond to
 the representation $\rho_{\ap}\oplus\rho_{\beta+\delta}$ and 
 $\rho_{\ap+\delta}\oplus\rho_{\beta}$, respectively. The mod $N$ integers
  $\ap,\beta,\delta$ are arbitrary. The reason why we restrict to this form 
  is because we want to maintain the ${\ca{N}}=2$ supersymmetry 
  in the presence 
 of the boundary perturbation. Indeed if we assume this form, we can read 
 off from eq.(\ref{t4}) the $g$-action on boundary fermionic superfields as
 follows
\ba
g\ :\ {\bf \Gamma}_1\ \to\ e^{\f{2\pi i}{N}(\ap-\beta)}\cdot
{\bf \Gamma}_1,\ \ \ \
{\bf \Gamma}_2\ \to\ e^{\f{2\pi i}{N}\delta}\cdot{\bf \Gamma}_2\ .
\ea
Further we can assume that $\beta\geq\ap$ and $\delta\geq 0$ without losing
generality. 

The holomorphic tachyon fields are classified into the form 
eq.(\ref{t2}) with $a=b=0$ and in addition they should be 
${\bf Z}_N$-invariant.
 Here we are interested in the generation of only D0-branes and thus we
 assume $q=r=0$ below\footnote{As we saw in section 2, some $\ddbt$ systems
  will be generated for $qr\neq 0$. Note that we cannot deform this as in
  (\ref{t2}) because of the orbifold projection.}.

Then the tachyon fields are classified into the following form
\ba
T_1(z_1,z_2)=a_{q}\cdot (z_1)^{\beta-\ap+Nq},\ \ \ 
T_2(z_1,z_2)=b_{r}\cdot (z_2)^{\delta+Nr}. \label{t22}
\ea

In the same way as the codimension two case, we can conclude that 
$(\beta-\ap+Nq)(\delta+Nr)$ fractional D0-branes will be generated.
 This can also be regarded as purely fractional D0-branes and 
 bulk D0-branes.
 The number of the former is given by $(\beta-\ap)\delta$ mod $N$. 
 In particular if one sets $\delta=0$ or $\beta=\ap$, then we obtain only
 bulk D0-branes. This is consistent with the fact that the two branes 
 and the
 two antibranes have identical type for these cases. The more detailed 
 argument which uses twisted RR-charges will be discussed in the next 
 section.

\subsection{Tachyon Condensation on Orbifolds and Twisted RR-charges}
\hspace{5mm}
Here we discuss the previous examples of tachyon condensation on the 
orbifolds
 from somewhat different viewpoint: we pay attention to the twisted 
 RR-charges
 in the orbifold theories. 
 
 Generally, an orbifold theory in the closed string 
 sector \cite{orbifold} consists of a untwisted-sector and 
 twisted-sectors. Our orbifold 
${\mb{C}}^2/{\bf Z}_N$ possesses $(N-1)$ twisted-sectors, which are twisted 
by
 $g,g^2,\ddd,g^{N-1}$. In each of the twisted NSNS-sectors there are
  four massless scalars and these correspond to the moduli of 
  hyper K\"ahler geometry. On the other hand, in each of the twisted 
  RR-sectors there is one vector field for Type IIA theory. The RR-charges 
  for these vector fields are called twisted RR-charges. 
  
These charges are carried by D-branes 
which do not belong to the regular representation. In other words, these 
represent the geometrical information that the branes are wrapped on some
 non-trivial 2-cycles in the ALE space. Therefore we argue that the twisted 
 RR-charges should be conserved during the tachyon condensation\footnote{
 Similar conservation law for D-branes in NS5-brane background 
 was recently discussed in \cite{HiNoSu}.}. Our example
 is consistent with this claim as we will see below. Note that this claim 
 is in strikingly contrast with the fact that for the untwisted 
 (or equally bulk) 
 RR-charge the generation of lower dimensional D-brane charges 
 does indeed occur. 
 This is due to the non-compactness of the orbifold. If one consider the 
orbifolded torus, then the untwisted charges should be conserved. 
Indeed the results 
from the description of tachyon condensation as the marginal deformation 
were 
obtained in ${\bf Z}_2$-orbifolded torus \cite{Sen9812,MaSe,NaTaUe} 
and the results are 
consistent with this.

 To see this more generally, let us remember the RR-coupling formula 
 (\ref{wz}).
 For compact space,
 as shown in \cite{Qu} the Chern character of the superconnection does not
 change in cohomology if we shift the value of the tachyon fields 
 continuously.
 This also supports the above arguments.

Now let us return to our examples in ${\mb{C}}^2/{\bf Z}_N$ . 
In principle the calculations of 
twisted RR-charges are possible in BSFT, but
 the determination of the normalizations is not so easy. Therefore we 
 calculate the charges in the boundary state formalism. For boundary 
 states in orbifold theories see for example 
 \cite{OoOzYi,HuIeNuSc,Sen98n,BiCrRo,Br,DiGo,Ta,BiCrRo2,BeDi}. 
 This formalism is 
useful to know couplings with various fields in closed string sector
because the boundary state is the description of a D-brane from the 
viewpoint
 of closed string theory. The detailed 
computations are shown in the appendix A and here we will discuss the 
results. 
 
The outline of the determination is as follows. First we can find the 
boundary 
state which represents a $\ap$-type D$p$-brane so as to satisfy the Cardy's 
condition 
\cite{Ca}
. Then the total boundary state is given by 
\ba
|\mbox{D}p(\ap)\lb=\sum_{k=0}^{N-1}e^{\f{2\pi ik\ap}{N}}|T^{(k)}\lb.
\ea
Here we defined the boundary states for untwisted sector 
$|T^{(0)}\lb=|U\lb$ and $k$-th twisted sectors $|T^{(k)}\lb$ as follows
\ba
|U\lb &=&\f{T_{p}}{2}(|U\lb _{NSNS}+|U\lb _{RR}), \no
|T^{(k)}\lb &=&\f{T_{p}'}{2}(|T^{(k)}\lb _{NSNS}+|T^{(k)}\lb _{RR}),
\label{dut}
\ea 
where the two normalization $T_{p}$ and $T_{p}'$ can be computed as in 
eq.(\ref{norm}). Next note that in the low energy limit the boundary state 
for each sector is proportional to a massless state in the sector. 
Thus we can read off the 
coupling to the massless field from the coefficient of the 
boundary state for each sector \cite{DiFrPeScLeRu,DiGo,Ta,BiCrRo2,BeDi}.

In this way we can compute the twisted RR-charges and the result is
  as follows for the $k$-th twisted RR-charge $Q^{(k)}_{\ap,p}$ of 
a $\ap$-type D$p$-brane 
\ba
Q^{(k)}_{\ap,p}=\f{1}{N}\cdot e^{\f{2\pi i}{N}\ap k}\cdot 
\left(2\sin\f{\pi k}{N}
\right)^{1-\f{p}{2}}\cdot2^{2}\pi^{\f32}(\al)^{\f12}.
\ea
Note that the above method cannot determine the phase factors which
 do not depend on $\ap$.

Then let us discuss the twisted RR-charges before and after the 
tachyon condensation. First consider the generation of the  
D($p$-2) brane charge
from a $\ddbp$ by the tachyon field (\ref{t11}). 
The original $\ddbp\ \ (p=2,4)$ has
 the $k$-th twisted RR-charge $(Q^{(k)}_{\ap,p}-Q^{(k)}_{\beta,p})$. Without 
losing generality we can assume $\beta \geq \ap$. If one 
notes the following elementary formula
\ba
\left(e^{\f{2\pi ik}{N}\ap}-
e^{\f{2\pi ik}{N}\beta}\right)=(-i)\cdot e^{\f{i\pi k}{N}}\cdot 
2\sin(\f{\pi k}{N})\cdot\left(e^{\f{2\pi ik}{N}\ap}+\ddd+
e^{\f{2\pi ik}{N}(\beta-1)}\right),
\ea
then one obtains 
\ba
Q^{(k)}_{\ap,p}-Q^{(k)}_{\beta,p}=(-i)\cdot e^{\f{\pi ik}{N}}\cdot
\sum_{\mu=\ap}^{\beta-1}Q^{(k)}_{\mu,p-2}\ .
\ea
This shows that {\it the final state after the tachyon condensation on a
$\ddbt$
should be fractional D0-branes of type $\{\ap,\ap+1,\ddd,\beta-1\}$} with 
some bulk D0-branes \footnote{The extra phase $(-i)\cdot e^{\f{\pi ik}{N}}$ 
can be canceled by the phase factor which cannot be determined from the 
calculations in the 
appendix A because it does not depend on $\ap$. The origin
 of $e^{\f{\pi ik}{N}}$ is easy to understand. If one considers the D$p$-
 D($p$-2) open string, then the $g$-action on the fermionic zero mode 
 generates the factor $e^{\pm\f{i\pi}{N}}$. 
Thus one must project the open string as 
$g=e^{\f{2\pi i}{N}(\f12+\beta-\ap)}$.}. This is consistent with the 
results in the previous 
 subsection that the 
 final state consists of $q$ bulk D0-branes and $(\beta-\ap)$ 
 fractional D0-branes. Combining this with the above arguments we 
 can determine the final state completely. 
 
 For $p$=4, one can also consider
 more general tachyon field (\ref{ddb41}). These will produce the 
 intersecting D2-brane system as mentioned in the previous subsection.
 Then we can find that the twisted charges are conserved if the charges 
 of $r$ D2-branes on $z_1=0$ do cancel those of $z_2=0$. 
 Note also that this configuration is BPS.

Next we turn to the generation of codimension four D-brane charges from
 the tachyon fields (\ref{t22}). In the same way as before we obtains 
 the following formula
\ba
Q^{(k)}_{\ap,4}+Q^{(k)}_{\beta+\delta,4}-Q^{(k)}_{\beta,4}-
Q^{(k)}_{\ap+\delta,4}=(i)^2\cdot\sum_{\mu=\ap+1}^{\beta}
\sum_{\nu=0}^{\delta-1}
Q^{(k)}_{\mu+\nu,0}\ , \label{qq}
\ea
where we assumed $\beta \geq\ap$ and $\delta\geq 0$. This decomposition rule
 is again consistent with the result in the previous subsection. Thus we can
 conclude that {\it after the
tachyon condensation there are $(\beta-\ap)\delta$ fractional D0-branes and 
their types are given by the above formula}.

In both examples of generating D0-branes 
if we shift the K\"ahler moduli and blow up the orbifold
 singularities, then we get (mutually) BPS D2-branes which are wrapped 
 on the corresponding holomorphic 2-cycles in ALE spaces. Since the 
 K\"ahler structure is
 independent from the complex structure, these holomorphic 2-cycles can be 
 ``defined" by the the equations $T_{i}=0$ in the $A_{(N-1)}$-type 
 hypersurface
\ba
XY=Z^N,\ \ \ (X=z_1^N,\ Y= z_2^N,\ Z=z_1z_2).
\ea
It will be also interesting to discuss the relation between the 
shift of complex structure and the corresponding tachyon field for
 these examples and we will leave this for future problem.

\subsection{Some Comments on Generalizations}
\hspace{5mm}

Before we close this section, let us comment on some 
generalizations of our results. First it is easy to see that 
the generalizations for 
higher dimensional ${\bf Z}_N$-orbifolds ${\mb{C}}^n/{\bf Z}_N$ $(n\geq 3)$ 
are straightforward since the above arguments largely depend 
on the algebraic
 properties of the discrete groups ${\bf Z}_N$.
 
 On the other hand, for other types of orbifolds ${\mb{C}}^n/{\bf \Gamma}$ 
 the results will be 
 non-trivial. Here we do not investigate these further, but it may be 
 natural 
 to conjecture that the following general 
 relation will hold for each $g\in {\bf \Gamma}$ with a coefficient $C(g)$
\ba
\sum_{i=1}^{2^{n-1}}\chi_{\ap_i}(g)-\sum_{i=1}^{2^{n-1}}\chi_{\beta_i}(g)
=C(g)\cdot \sum_{\delta\in \Delta}
\chi_{\delta}(g), \label{cf}
\ea
where $\chi_{\ap}(g)$ is the character of $g$ for the irreducible 
representation $\ap$ and $\Delta$ denotes a certain subset of irreducible 
representations which depends on $\ap_i,\beta_i$. 
Note that if we return to the ${\mb{C}}^2/{\bf Z}_N$ examples, then 
the character is given by $\chi_{\ap}(g^k)=\exp(2\pi i \ap k/N)$ and
 the relation eq.(\ref{cf}) is equivalent to eq.(\ref{qq}). The coefficient 
 $C(g)$ will be due to a phase factor and due to the trace over
 the zeromodes as in (\ref{zts}). 
 
\section{Conclusions}
\setcounter{equation}{0}
\hspace{5mm}

In this paper we have discussed the boundary string field theory description 
of tachyon condensation with world-sheet ${\ca{N}}=2$ supersymmetry. This
extended supersymmetry generally requires that the tachyon field should be 
holomorphic \cite{Ho}. Therefore it is natural to believe that this 
constraint is related to the spacetime supersymmetry of final states 
after the 
 tachyon condensation. We have investigated this issue in two examples.

First we have
 considered brane-antibrane systems in flat space and discuss the 
 generalization
  of Atiyah-Bott-Shapiro configuration. In the arguments of these the 
  RR-coupling formula for brane-antibrane systems also played an 
  important role.
  As a result, we obtained only BPS configurations from the minimal number 
  of 
  brane-antibrane pairs.
  
Next we have investigated tachyon condensation on ${\bf Z}_{N}$-orbifolds
 mainly in four dimension.
This is one of the simplest examples in curved spaces and 
most of our arguments can be performed algebraically. 
In this example we have seen that
 holomorphic tachyon fields generate various BPS fractional D-branes 
 which are wrapped on various holomorphic cycles. The conservation law of
 various twisted RR-charges was used to identify the final states.

Finally let us mention some future directions. If one wants to see the 
generation of lower dimensional D-branes explicitly, it will be useful to
 construct the (off-shell) boundary states during the tachyon condensation 
 in the same way as in \cite{FrGaLeSt,matsuo,NaTaUe,deA}.
 This will make more clear 
 the generation of fractional D-branes from brane-antibrane systems.
 
 In particular for BPS D-branes on the four dimensional orbifolds
  (or K3 surface), 
 the world-sheet
 ${\ca{N}}=4$ superconformal symmetry is realized \cite{OoOzYi}. 
 Thus it is intriguing to construct ${\ca{N}}=4$ boundary interactions
  and discuss tachyon condensation in BSFT.
 
  As mentioned in section 3.3, it will also interesting to investigate 
  other examples of orbifolds 
  because the consideration of tachyon condensation seems to imply 
  non-trivial relations among the characters of irreducible 
  representations. 
  
  We hope 
  to return to these issues in future work.

\bigskip

\begin{center}
\noindent{\large \bf Acknowledgments}
\end{center}
I am very grateful to T.Eguchi for encouragement and valuable advice.
I also thank 
Y.Hikida, Y.Matsuo, M.Naka, M.Nozaki, K.Ohmori, 
Y.Sugawara, S.Terashima and T.Uesugi for useful
 discussions. This work is supported by JSPS Research Fellowships for Young 
 Scientists.
\appendix
\setcounter{equation}{0}
\section{Detailed Boundary State Computations}
\hspace{5mm}
Here we compute the cylinder amplitudes of open strings between fractional 
D$p$-branes ($p=0,2,4$) on the orbifold ${\mb{C}}^2/{\bf Z}_N$ in order to 
get correct normalizations of the boundary states. Similar calculations for 
$p=0$ or for ${\bf Z}_2$-orbifolds have 
been performed in various papers, for example 
\cite{HuIeNuSc,Sen98n,BiCrRo,Br,DiGo,Ta,BiCrRo2,BeDi} (see also \cite{GiJo}).
Let us first summarize our conventions.

\begin{flushleft}
{\bf Conventions for Open String}
\end{flushleft}

We define the open string Hamiltonian of world-sheet theory as
\ba
H_{o}=\pi(\al p^\mu p_\mu+N_{o}+a),
\ea
where $p^\mu$ is the momentum and  $N_{o}\in {\bf Z}$ is the contributions 
from oscillators; $a$ denotes the
zero energy 
\ba
a=-\f{1}{2}\ \  \mbox{(for NS-sector)},\ \ \  a=0\ \  \mbox{(for R-sector)}.
\ea

The moduli of the cylinder is written by $t$ and we define $q=e^{2\pi i\tau}$ 
as
\ba
q=e^{2\pi i\tau}\equiv e^{-2\pi t}.
\ea

The one-loop amplitude $Z_{open}$ of open string between a $\ap$-type 
D$p$-brane and $\beta$-type D$p$-brane can be written as
\ba
Z_{open}=\f{1}{N}\sum_{k=0}^{N-1}e^{i\f{2\pi}{N}(\ap-\beta)k}\ Z^{(k)}_{open},
\label{ope}
\ea
where $Z^{(k)}_{open}$ is defined by
\ba
Z^{(k)}_{open}=2\int_{0}^{\infty}\f{dt}{2t} \mbox{Tr}_{NS-R}\left
[\ g^k \f{1+(-1)^F}{2}
e^{-2H_{o}t}\ \right].
\ea

This means the ${\bf Z}_N$-projection into the states which satisfy 
$g=e^{i\f{2\pi}{N}(\beta-\ap)}$.

Next let us consider the bosonic zeromodes along ${\mb{C}}^2/{\bf Z}_N$ 
direction. The traces over these zeromodes become
\ba
\mbox{Tr}(1)=\mbox{V}_{p}\cdot \int (\f{dk}{2\pi})^p, \ \ \ 
\mbox{Tr}(g^k)=\f{1}{(2\sin(\f{\pi k}{N}))^p}, \label{zts}
\ea
where $\mbox{V}_{p}$ denotes the volume of a D$p$-brane before the 
${\bf Z}_N$-projection. The second equation follows from the 
calculation \cite{GiJo}
\ba
\int (dz)^2\ \la z|g^k|z\lb
=\f{1}{2\sin(\f{\pi k}{N})}\ \ \ \  \left(\la z|z'\lb\equiv \delta^2(z-z')
\right).
\ea

Then we turn to the fermionic zeromodes in the R-sector 
along ${\mb{C}}^2/{\bf Z}_N$ direction. The action of $g$ on these 
is defined as follows:
\ba
g\ |s_1,s_2\lb=e^{\f{2\pi i}{N}(s_1-s_2)}|s_1,s_2\lb,
\ea
where $s_1,s_2\in \{\pm\f12\}$ denote the spins of the spacetime fermions. 
From
 this one can obtain the zeromode trace in R-sector as
\ba
\mbox{Tr}_{R}(g^k)=e^{\f{2\pi ik}{N}}+e^{-\f{2\pi ik}{N}}+2=4\cos^2
\left(\f{\pi k}{N}\right).
\ea
Below we use the trace $\mbox{tr}$ over only oscillators 
(not the bosonic and fermionic zeromodes).

\begin{flushleft}
{\bf Formulae of $\theta$-functions}
\end{flushleft}

Here we summarize the formulae of $\theta$-functions. First we define the 
following $\theta$-functions:
\ba
\eta(\tau)&=&q^{\f{1}{24}}\prod_{n=1}^{\infty}(1-q^n),\no
\theta_{1}(\nu,\tau)&=&2q^{\f18}\sin(\pi\nu)\prod_{n=1}^{\infty}(1-q^n)
(1-e^{2i\pi\nu}q^{n})(1-e^{-2i\pi\nu}q^{n}),\no
\theta_{2}(\nu,\tau)&=&2q^{\f18}\cos(\pi\nu)\prod_{n=1}^{\infty}(1-q^n)
(1+e^{2i\pi\nu}q^{n})(1+e^{-2i\pi\nu}q^{n}),\no
\theta_{3}(\nu,\tau)&=&\prod_{n=1}^{\infty}(1-q^n)
(1+e^{2i\pi\nu}q^{n-\f12})(1+e^{-2i\pi\nu}q^{n-\f12}),\no
\theta_{4}(\nu,\tau)&=&\prod_{n=1}^{\infty}(1-q^n)
(1-e^{2i\pi\nu}q^{n-\f12})(1-e^{-2i\pi\nu}q^{n-\f12}),
\ea
where we have defined $q=e^{2i\pi\tau}$.

Then the modular transformations are given as follows
\ba
\eta(\tau)&=&(-i\tau)^{-\f12}\eta(-1/\tau),\ \ \theta_{1}(\nu,\tau)
=i(-i\tau)^{-\f12}e^{-\pi i\f{\nu^2}{\tau}}\theta_{1}
(\nu/\tau,-1/\tau), \no
\theta_{2}(\nu,\tau)&=&(-i\tau)^{-\f12}e^{-\pi i\f{\nu^2}{\tau}}
\theta_{4}(\nu/\tau,-1/\tau), \ \ \theta_{3}(\nu,\tau)
=(-i\tau)^{-\f12}e^{-\pi i\f{\nu^2}{\tau}}\theta_{3}(\nu/\tau,-1/\tau)
, \no\theta_{4}(\nu,\tau)&=&(-i\tau)^{-\f12}e^{-\pi i\f{\nu^2}{\tau}}
\theta_{2}(\nu/\tau,-1/\tau).
\ea

\begin{flushleft}
{\bf Open String Cylinder Amplitudes}
\end{flushleft}

Let us compute the open string cylinder amplitudes $Z^{(k)}_{open}$. 
We only consider the two coincident D$p$-branes.

For 
the untwisted part $k=0$, we obtain
\ba
Z^{(0)}_{open}&=&2\mbox{V}_{p+1}\ \int_{0}^{\infty}\f{dt}{2t}\ 
(8\pi^2\al t)^{-\f{p+1}{2}}\cdot\f{\theta_{3}(0,it)^4-\theta_{4}(0,it)^4-
\theta_{2}(0,it)^4}{2\eta(it)^{12}} \\
&=&2^{-\f{3p+5}{2}}\pi^{\f{-3p+5}{2}}(\al)^{-\f{p+1}{2}}\mbox{V}_{p+1}\cdot
\int ds\ s^{\f{p-9}{2}}\cdot
\f{\theta_{3}(0,is/\pi)^4-\theta_{2}(0,is/\pi)^4-
\theta_{4}(0,is/\pi)^4}{2\eta(is/\pi)^{12}}, \nonumber
\ea
where $\mbox{V}_{p+1}$ is equal to $\mbox{V}_{p}$ times the ``volume" 
V$_1$ of 
time-like direction. Note that in the last expression we have performed the 
modular transformation. For the $k$-th twisted parts, the result is
\ba
Z^{(k)}_{open}&=&\f{2\mbox{V}_1}{(2\sin\f{\pi k}{N})^p}
\int_{0}^{\infty}\f{dt}{2t}
(8\pi^2\al t)^{-\f{1}{2}}\f{(2\sin\f{\pi k}{N})^2}
{\theta_{1}(\f{k}{N},it)^2}\cdot \f{1}{2\eta(it)^{6}}\\
& &\ \times [\theta_{3}(0,it)^2\theta_{3}(k/N,it)^2-
\theta_{4}(0,it)^2\theta_{4}(k/N,it)^2-
\theta_{2}(0,it)^2\theta_{2}(k/N,it)^2]\no
&=&2^{-\f52}\pi^{\f12}\al^{-\f12}\mbox{V}_{1}\f{(2\sin\f{\pi k}{N})^{2-p}}
{(i)^2}\int ds\ s^{-\f52}\f{(2\sin\f{\pi k}{N})^2}{\theta_{1}(\f{k}{N},it)^2}
\cdot\f{1}{\eta(is/\pi)^{6}
\theta_{1}(\nu_{k},is/\pi)^2}\no 
& &\times [\theta_{3}(0,is/\pi)^2\theta_{3}(\nu_{k},is/\pi)^2-
\theta_{4}(0,is/\pi)^2 \theta_{4}(\nu_{k},is/\pi)^2-
\theta_{2}(0,is/\pi)^2 \theta_{2}(\nu_{k},is/\pi)^2] \nonumber
,
\ea
where we have defined $\nu_{k}=-iks/N\pi$.

Then let us compare these results with those from the boundary state 
calculations. Before that we summarize the conventions. We use the 
light-cone gauge in NS-R formulation \cite{BeGa2} and closely follow 
the normalization in \cite{DiFrPeScLeRu}.

\begin{flushleft}
{\bf Conventions for Boundary State}
\end{flushleft}

The closed string Hamiltonian is defined by
\ba
H_{c}=\pi\al k^\mu k_\mu+2\pi(N_{L}+N_{R})+4\pi a,
\ea
where $N_{L}$ and $N_{R}$ are the contributions from left-moving and 
right-moving 
oscillators; $a$ denotes the
zero energy 
\ba
a=-\f{1}{2}+\f{k}{N}\ \  \mbox{(for NSNS-sector)},\ \ \  a=0\ \  
\mbox{(for RR-sector)}.
\ea
Note also that the momentum $k^\mu$ in twisted sectors is always zero along
 the orbifold direction ${\mb{C}}^2/{\bf Z}_N$.

Further one can define the propagator $\Delta$ as
\ba
\Delta=\f{\al}{2}\int ds\  e^{-\f{1}{2\pi}sH_{c}}.
\ea

The boundary state for the untwisted-sector and $k=1,2,\ddd,(N-1)$ -th 
twisted-sectors are given by
\ba
|U\lb &=&\f{T_{p}}{2}(|U\lb _{NSNS}+|U\lb _{RR}), \no
|T^{(k)}\lb &=&\f{T_{p}'}{2}(|T^{(k)}\lb _{NSNS}+|T^{(k)}\lb _{RR}), 
\label{ut}
\ea 
where the constants $T_{p},T_{p}'$ represent 
the tension and charges of the D-brane
and will be determined later. 
We have defined $|U\lb_{sector}$ and $|T^{(k)}\lb_{sector}$ as
\ba
|U\lb _{NSNS}&=&\f{1}{2}\int (\f{dk}{2\pi})^{9-p}(|U,+,k^a\lb_{NSNS}
-|U,-,k^a\lb_{NSNS}), \\
|U\lb _{RR}&=&2\int (\f{dk}{2\pi})^{9-p}(|U,+,k^a\lb_{RR}
+|U,-,k^a\lb_{RR}), \\
|T^{(k)}\lb _{NSNS}&=&\f{1}{2}\int (\f{dk}{2\pi})^{5}
(|T^{(k)},+,k^i\lb_{NSNS}-|T^{(k)},-,k^i\lb_{NSNS}), \label{bs} \\
|T^{(k)}\lb _{NSNS}&=&\int (\f{dk}{2\pi})^{5}
(|T^{(k)},+,k^i\lb_{RR}+|T^{(k)},-,k^i\lb_{RR}), 
\ea
where $k^a$ and $k^i$ are momenta of the D$p$-brane in the 
untwisted and twisted sectors, respectively. If we regard $x^6,\ddd,x^9$ 
as the coordinates of ${\mb{C}}^2/{\bf Z}_N$, then we can take 
$a=1,\ddd,9-p$ and
 $i=1,2,\ddd,5$. The explicit forms of $|U,\pm,k^a\lb_{sector},
 \ |T^{(k)},\pm,k^i\lb_{sector}$ are determined by the requirement that
 they should satisfy the desirable boundary conditions. These conditions
 are solved by elementary calculations and the explicit forms are given 
 by ``coherent states'' of left 
 and right-moving oscillators. Here we show the explicit expression only 
 for $p=0$ in NSNS
 -sector as follows (we assume $k<N/2$ for simplicity of the notation 
 and we 
define $T^{(0)}=U$)

\ba
 |T^{(k)},\ep,\vec{k}\lb_{NSNS}&=&\exp\left[\sum_{n=1}^{\infty}
 \left(\f{1}{n}
 \sum_{\mu=2}^{5}
\ap^\mu_{-n}\tap^\mu_{-n}\right)+i\ep\sum_{r > 0}\left
(\sum_{\mu=2}^{5}\p^\mu_{-r}\bp^\mu_{-r}\right)\right] 
\nonumber  \\
  & &\times \exp\biggl[\sum_{n=0}^{\infty}\left(\f{1}{n+\fkN}\ap_{-n-\fkN}
  \tbap_{-n-\fkN}\right)+\sum_{n=1}^{\infty}\left(\f{1}{n-\fkN}
  \tap_{-n+\fkN}
  \bap_{-n+\fkN}\right) \nonumber  \\
  & &+\sum_{n=0}^{\infty}\left(\f{1}{n-\fkN}\beta_{-n+\fkN}\tbbe_{-n+\fkN}
  \right)+
  \sum_{n=1}^{\infty}\left(\f{1}{n+\fkN}\tbe_{-n-\fkN}\bbe_{-n-\fkN}
  \right)  
  \nonumber  \\
  & &+i\ep\left(\sum_{r > 0}\et_{-r-\fkN}\tbet_{-r-\fkN}+\sum_{r > 0}
  \tet_{-r+\fkN}\bet_{-r+\fkN}\right)            \nonumber  \\
  & &+i\ep\left(\sum_{r > 0}\xi_{-r+\fkN}\tbxi_{-r+\fkN}+\sum_{r > 0}
  \txi_{-r-\fkN}\bxi_{-r-\fkN}\right)\biggr]|T^{(k)},\ep,\vec{k}
  \lb^{(0)}_{NSNS}, 
  \label{tkkn} 
\ea
where we defined the zeromode as $|T^{(k)},\ep,\vec{k}\lb^{(0)}_{NSNS}$.
The oscillators $(\ap^\mu,\tap^\mu)$ and $(\p^\mu,\bp^\mu)$ are
 for bosonic fields $(X_{L}^\mu,X_{R}^\mu)$ and for fermionic fields 
 $(\psi_{L}^\mu,\psi_{R}^\mu)$ on the world-sheet; $(\ap,
 \tap,\beta,\tbe)$ denote the oscillators for 
 $(Z_{L}^1,Z_{R}^1,Z_{L}^2,Z_{R}^2)$ and $(\et,\tet,\xi,\txi)$ 
 are their superpartners. They
 follow the canonical (anti)commutation relations
\ba
&&[\bap_{m+\fkN}\ ,\ \ap_{n-\fkN}]=(m+k/N)\delta_{m,-n}\ \ , \ 
\ [\bbe_{m-\fkN}\ ,\ \beta_{n+\fkN}]=(m-k/N)\delta_{m,-n} \nonumber \\
&&\{\eta_{r-\fkN}\ ,\ \bet_{r+\fkN}\}=\delta_{r+s}\ \  \ \ ,\ \ \  \ 
\ \{\xi_{r+\fkN}\ ,\ \bxi_{r-\fkN}\}=\delta_{r+s} 
\ea

The expressions for the others are also written almost in the same form 
as (\ref{tkkn}). For more details we recommend the readers to refer to 
\cite{BiCrRo,DiGo,Ta,BiCrRo2}, for example.

We also comment that the above definition
 (\ref{bs}) does not work for $k=\f{N}{2}$ because there are extra fermionic
 zeromodes in twisted NSNS-sector along the orbifold direction. In this case
 one should change the factor in front of R.H.S. of (\ref{bs}) into $1$ and 
 the sign in the middle of (\ref{bs}) into $+$.

Next the zeromodes are normalized as follows: for the untwisted and twisted 
sectors
\ba
\la k^a|k'^a \lb^{(0)} =\mbox{V}_{p+1}(2\pi)^{9-p}\delta^{9-p}(k^a-k'^a),
\ \ \ \la k^i|k'^i\lb^{(0)}=\mbox{V}_{1}(2\pi)^{5}\delta^{5}(k^i-k'^i). 
\label{zn}
\ea

Finally we get the total boundary state $|\mbox{D}p(\ap)\lb$ 
which describes a 
$\ap$-type D$p$-brane
 as follows:
\ba
|\mbox{D}p(\ap)\lb=\sum_{k=0}^{N-1}e^{\f{2\pi ik\ap}{N}}|T^{(k)}\lb. 
\label{bspp}
\ea
The phase factors $e^{\f{2\pi ik\ap}{N}}$ 
are inserted in order to be consistent with the open string calculations. 
These are proportional to the charges in twisted-sectors.

\begin{flushleft}
{\bf Open-Closed duality}
\end{flushleft}

As argued by Cardy \cite{Ca}, the one-loop amplitude of open string should
be equal to the tree level amplitude between two boundary states in
closed string. This requirement is called Cardy's condition and often gives
a crucial consistency condition of D-branes. In our case, we can write this
requirement as follows:
\ba
Z_{open}=\la\mbox{D}p(\beta)|\Delta|\mbox{D}p(\ap)\lb.
\ea
Then comparing this with the eq.(\ref{ope}) and using (\ref{bspp}), we obtain
\ba
\f{1}{N}Z^{(k)}_{open}&=&\la T^{(k)}|\Delta|T^{(k)}\lb. \no
\ea

\begin{flushleft}
{\bf Boundary State Calculations and Determination of the Normalization}
\end{flushleft}

Now let us compute the cylinder amplitude in the boundary state formalism.
The result for untwisted-sector is given by 
\ba
&&\la U|\Delta|U\lb \\
& &=\f{V_{p+1}T_p^2\al}{16}\int \left(\f{dk}{2\pi}\right)^{9-p}
ds \ e^{-\f{1}{2}\al k^2 s}\cdot
\f{\theta_{3}(0,is/\pi)^4
-\theta_{2}(0,is/\pi)^4-\theta_{4}(0,is/\pi)^4}{\eta(is/\pi)^{12}}.\nonumber
\ea
For $k$-th twisted-sectors we obtain
\ba
\la T^{(k)}|\Delta|T^{(k)}\lb&=&\f{V_{1}(T'_p)^2\al}{16}
\int (\f{dk}{2\pi})^{5} ds e^{-\f{1}{2}\al k^2 s}\
\eta(is/\pi)^{-6}\left((-i)\theta_{1}(\nu_{k},is/\pi)\right)^{-2}\no
&\times&\biggl[\theta_{3}(0,is/\pi)^2\theta_{3}(\nu_{k},is/\pi)^2-
\theta_{4}(0,is/\pi)^2 \theta_{4}(\nu_{k},is/\pi)^2\no
& &-\theta_{2}(0,is/\pi)^2 \theta_{2}(\nu_{k},is/\pi)^2\biggr].
\ea

Then after we perform the integration in the above equations, we
can determine the normalizations $T_p,\ T'_p$ from the Cardy's condition:
\ba
T_{p}&=&\f{1}{\s{N}}\cdot 2^{3-p}\pi^{\f72-p}(\al)^{\f{3-p}{2}}, \no
T'_{p}&=&\f{1}{\s{N}}\cdot2^{2}\pi^{\f32}(\al)^{\f12}\cdot 
\left(2\sin\f{\pi k}{N}
\right)^{1-\f{p}{2}}. \label{norm}
\ea

\begin{flushleft}
{\bf Tension and Charges}
\end{flushleft}

Finally let us determine the tension $T_{Dp}$ and $k$-th twisted RR-charges 
$Q^{(k)}_{\ap,p}$ of a $\ap$-type
 D$p$-brane. Generally, one can compute a coupling with a closed string 
 field
from the overlap of the boundary state with the 
 corresponding vertex operator as discussed in \cite{DiFrPeScLeRu}. Therefore
  the tension and twisted RR-charges of our example can also be read off 
  from the boundary state 
  $|\mbox{D}p(\ap)\lb$ (\ref{bspp}) as follows
\ba
T_{D0}&=&\f{T_{0}}{\s{N}},\ \ \ T_{D2}=\s{N}T_{2},\ \ \ T_{D4}=\s{N}T_{4}
,\no
Q^{(k)}_{\ap,p}&=&\f{1}{\s{N}}\cdot e^{\f{2\pi i}{N}\ap k}\cdot T'_{p},
\ea
where the factor $\f{1}{\s{N}}$ is needed for the correct normalization of 
untwisted fields \cite{BiCrRo2}; the different coefficients of the tensions 
for  $p=2,4$ are due to the facts that 
the volume factor $\mbox{V}_{p+1}$ in (\ref{zn}) should be divided by $N$
 in physical context. Then it is obvious that the tension of a $\ap$-type 
 D0-brane is $\f{1}{N}$ times that of an ordinary D0-brane in flat space. 
 On the other hand,
 for a $\ap$-type D2 or D4-brane the tension is the same as that of 
 a ordinary D-brane. Some aspects of twisted RR-charges 
 were discussed 
 in section 3. Note that the D-brane also has a untwisted charge and 
 twisted NSNS charges, which are proportional to the tension and the 
 twisted RR-charges, respectively.


\begin{thebibliography}{99}
\small
\baselineskip=15pt
\bibitem{BA1}
M.~B.~Green,
``Pointlike states for type 2b superstrings,''
Phys.\ Lett.\ B {\bf 329} (1994) 435
[hep-th/9403040].

\bibitem{BA2}
T.~Banks and L.~Susskind,
``Brane - Antibrane Forces,''
hep-th/9511194.

\bibitem{Sen985}
A.~Sen,
``Tachyon condensation on the brane antibrane system,''
JHEP{\bf 9808} (1998) 012
[hep-th/9805170].

\bibitem{BeGa}
O.~Bergman and M.~R.~Gaberdiel,
``Stable non-BPS D-particles,''
Phys.\ Lett.\ B {\bf 441} (1998) 133
[hep-th/9806155].

\bibitem{Sen994}
A.~Sen,
``Non-BPS states and branes in string theory,''
hep-th/9904207.

\bibitem{Hal}

K.~Bardakci, 
``Dual Models and Spontaneous Symmetry Breaking,''
Nucl.\ Phys.\ B {\bf 68} (1974) 331; 
``Spontaneous Symmetry Breakdown In The Standard Dual String Model,''
Nucl.\ Phys.\ B {\bf 133} (1978) 297.

K.~Bardakci and M.~B.~Halpern,
``Explicit Spontaneous Breakdown In A Dual Model,''
Phys.\ Rev.\ D {\bf 10} (1974) 4230; ``Explicit Spontaneous 
Breakdown In A Dual Model. 2. N Point Functions,''
Nucl.\ Phys.\ B {\bf 96} (1975) 285.

\bibitem{Witten}
E.~Witten,
``Noncommutative Geometry And String Field Theory,''
Nucl.\ Phys.\ B {\bf 268} (1986) 253.

\bibitem{Be}
N.~Berkovits,
``SuperPoincare invariant superstring field theory,''
Nucl.\ Phys.\ B {\bf 450} (1995) 90
[hep-th/9503099]; ``A new approach to superstring field theory,''
Fortsch.\ Phys.\ {\bf 48} (2000) 31
[hep-th/9912121].

\bibitem{Cubic}
V.~A.~Kostelecky and S.~Samuel,
``On A Nonperturbative Vacuum For The Open Bosonic String,''
Nucl.\ Phys.\ B {\bf 336} (1990) 263; ``The Static Tachyon Potential
 In The Open Bosonic String Theory,''
Phys.\ Lett.\ B {\bf 207} (1988) 169.

A.~Sen and B.~Zwiebach,
``Tachyon condensation in string field theory,''
JHEP{\bf 0003} (2000) 002
[hep-th/9912249].

N.~Berkovits, A.~Sen and B.~Zwiebach,
``Tachyon condensation in superstring field theory,''
Nucl.\ Phys.\ B {\bf 587} (2000) 147
[hep-th/0002211].

\bibitem{Oh}
K.~Ohmori,
``A Review on Tachyon Condensation in Open String Field Theories,''
hep-th/0102085.

\bibitem{Witten2}
E.~Witten,
``On background independent open string field theory,''
Phys.\ Rev.\ D {\bf 46} (1992) 5467
[hep-th/9208027]; ``Some computations in background independent 
off-shell string theory,''
Phys.\ Rev.\ D {\bf 47} (1993) 3405
[hep-th/9210065].

\bibitem{Sh}
S.~L.~Shatashvili,
``Comment on the background independent open string theory,''
Phys.\ Lett.\ B {\bf 311} (1993) 83
[hep-th/9303143]; ``On the problems with background independence
 in string theory,''
hep-th/9311177.

\bibitem{GeSh}
A.~A.~Gerasimov and S.~L.~Shatashvili,
``On exact tachyon potential in open string field theory,''
JHEP{\bf 0010} (2000) 034
[hep-th/0009103].

\bibitem{KuMaMo}
D.~Kutasov, M.~Marino and G.~Moore,
``Some exact results on tachyon condensation in string field theory,''
JHEP{\bf 0010} (2000) 045
[hep-th/0009148].

\bibitem{Ts}
E.~S.~Fradkin and A.~A.~Tseytlin,
``Nonlinear Electrodynamics From Quantized Strings,''
Phys.\ Lett.\ B {\bf 163} (1985) 123;

O.~D.~Andreev and A.~A.~Tseytlin,
``Partition Function Representation For The 
Open Superstring Effective Action: Cancellation 
Of Mobius Infinities And Derivative Corrections To
Born-Infeld Lagrangian,''
Nucl.\ Phys.\ B {\bf 311} (1988) 205;

A.~A.~Tseytlin,
``Sigma model approach to string theory effective 
actions with tachyons,''
hep-th/0011033.

\bibitem{KuMaMo2}
D.~Kutasov, M.~Marino and G.~Moore,
``Remarks on tachyon condensation in superstring field theory,''
hep-th/0010108.

\bibitem{Ho}
K.~Hori,
``Linear models of supersymmetric D-branes,''
hep-th/0012179.

\bibitem{KrLa}
P.~Kraus and F.~Larsen,
``Boundary string field theory of the D D-bar system,''
hep-th/0012198.

\bibitem{TaTeUe}
T.~Takayanagi, S.~Terashima and T.~Uesugi,
``Brane-antibrane action from boundary string field theory,''
hep-th/0012210.

\bibitem{Sen988}
A.~Sen,
``SO(32) spinors of type I and other solitons on brane-antibrane pair,''
JHEP{\bf 9809} (1998) 023
[hep-th/9808141].

\bibitem{FrGaLeSt}
M.~Frau, L.~Gallot, A.~Lerda and P.~Strigazzi,
``Stable non-BPS D-branes in type I string theory,''
Nucl.\ Phys.\ B {\bf 564} (2000) 60
[hep-th/9903123].

\bibitem{matsuo}
Y.~Matsuo,
``Tachyon condensation and boundary states in bosonic string,''
hep-th/0001044.

\bibitem{Sen9812}
A.~Sen,
``BPS D-branes on non-supersymmetric cycles,''
JHEP{\bf 9812} (1998) 021 [hep-th/9812031].

\bibitem{MaSe}
J.~Majumder and A.~Sen,
``'Blowing up' D-branes on non-supersymmetric cycles,''
JHEP{\bf 9909} (1999) 004
[hep-th/9906109]; ``Vortex pair creation on brane-antibrane pair 
via marginal deformation,'' JHEP{\bf 0006} (2000) 010 [hep-th/0003124];
``Non-BPS D-branes on a Calabi-Yau orbifold,''
JHEP{\bf 0009} (2000) 047
[hep-th/0007158].

\bibitem{NaTaUe}
M.~Naka, T.~Takayanagi and T.~Uesugi,
``Boundary state description of tachyon condensation,''
JHEP{\bf 0006} (2000) 007
[hep-th/0005114].

\bibitem{OzPaWa}
Y.~Oz, T.~Pantev and D.~Waldram,
``Brane-antibrane systems on Calabi-Yau spaces,''
hep-th/0009112.

\bibitem{Tatar}
R.~Tatar,
``A note on non-commutative field theory and stability of 
brane-antibrane  systems,''
hep-th/0009213.

\bibitem{HiNoTa}
Y.~Hikida, M.~Nozaki and T.~Takayanagi,
``Tachyon condensation on fuzzy sphere and noncommutative solitons,''
Nucl.\ Phys.\ B {\bf 595} (2001) 319
[hep-th/0008023].

\bibitem{HiNoSu}
Y.~Hikida, M.~Nozaki and Y.~Sugawara,
``Formation of spherical D2-brane from multiple D0-branes,''
hep-th/0101211.

\bibitem{NCS}
K.~Dasgupta, S.~Mukhi and G.~Rajesh,
``Noncommutative tachyons,''
JHEP{\bf 0006} (2000) 022
[hep-th/0005006];

J.~A.~Harvey, P.~Kraus, F.~Larsen and E.~J.~Martinec,
``D-branes and strings as non-commutative solitons,''
JHEP{\bf 0007} (2000) 042
[hep-th/0005031];

J.~A.~Harvey, P.~Kraus and F.~Larsen,
``Exact noncommutative solitons,''
JHEP{\bf 0012} (2000) 024
[hep-th/0010060].

\bibitem{BaKaMaTa}
I.~Bars, H.~Kajiura, Y.~Matsuo and T.~Takayanagi,
``Tachyon condensation on noncommutative torus,''
hep-th/0010101;

E.~M.~Sahraoui and E.~H.~Saidi,
``Solitons on compact and noncompact spaces in large noncommutativity,''
hep-th/0012259.

\bibitem{MaMo}
E.~J.~Martinec and G.~Moore,
``Noncommutative solitons on orbifolds,''
hep-th/0101199.

\bibitem{Do}
M.~R.~Douglas,
``Two lectures on D-geometry and noncommutative geometry,''
hep-th/9901146; ``Topics in D-geometry,''
Class.\ Quant.\ Grav.\ {\bf 17} (2000) 1057
[hep-th/9910170].

\bibitem{Witten3}
E.~Witten,
``D-branes and K-theory,''
JHEP{\bf 9812} (1998) 019
[hep-th/9810188].

\bibitem{HaKuMa}
J.~A.~Harvey, D.~Kutasov and E.~J.~Martinec,
``On the relevance of tachyons,''
hep-th/0003101.

\bibitem{BeBeSt}
K.~Becker, M.~Becker and A.~Strominger,
``Five-branes, membranes and nonperturbative string theory,''
Nucl.\ Phys.\ B {\bf 456} (1995) 130
[hep-th/9507158].

\bibitem{OoOzYi}
H.~Ooguri, Y.~Oz and Z.~Yin,
``D-branes on Calabi-Yau spaces and their mirrors,''
Nucl.\ Phys.\ B {\bf 477} (1996) 407
[hep-th/9606112].

\bibitem{ABS}
M.F. Atiyah, R. Bott and A. Shapiro, ``Clifford Modules'', Topology
 {\bf 3} (1964) 3.

\bibitem{Wa}
N.~P.~Warner,
``Supersymmetry in boundary integrable models,''
Nucl.\ Phys.\ B {\bf 450} (1995) 663
[hep-th/9506064].


\bibitem{KeWi}
C.~Kennedy and A.~Wilkins,
``Ramond-Ramond couplings on brane-antibrane systems,''
Phys.\ Lett.\ B {\bf 464} (1999) 206
[hep-th/9905195].

\bibitem{AlItOz}
M.~Alishahiha, H.~Ita and Y.~Oz,
``On superconnections and the tachyon effective action,''
hep-th/0012222.

\bibitem{Qu}D. Quillen, ``Superconnection and the Chern character,'' 
Topology {\bf 24} (1985) 89.

\bibitem{Te}
S.~Terashima,
``A construction of commutative d-branes from lower 
dimensional non-BPS  D-branes,''
hep-th/0101087.

\bibitem{orbifold}
L.~Dixon, J.~A.~Harvey, C.~Vafa and E.~Witten,
``Strings On Orbifolds,''
Nucl.\ Phys.\ B {\bf 261} (1985) 678;``Strings On Orbifolds. 2,''
Nucl.\ Phys.\ B {\bf 274} (1986) 285;

L.~Dixon, D.~Friedan, E.~Martinec and S.~Shenker,
``The Conformal Field Theory Of Orbifolds,''
Nucl.\ Phys.\ B {\bf 282} (1987) 13.

\bibitem{GC}
H.~Garcia-Compean,
``D-branes in orbifold singularities and equivariant K-theory,''
Nucl.\ Phys.\ B {\bf 557} (1999) 480
[hep-th/9812226].

\bibitem{DoMo}
M.~R.~Douglas and G.~Moore,
``D-branes, Quivers, and ALE Instantons,''
hep-th/9603167.

\bibitem{JoMy}
C.~V.~Johnson and R.~C.~Myers,
``Aspects of type IIB theory on ALE spaces,''
Phys.\ Rev.\ D {\bf 55} (1997) 6382
[hep-th/9610140].

\bibitem{EgGiHa}
T.~Eguchi, P.~B.~Gilkey and A.~J.~Hanson,
``Gravitation, Gauge Theories And Differential Geometry,''
Phys.\ Rept.\ {\bf 66} (1980) 213.

\bibitem{As}
P.~S.~Aspinwall,
``Enhanced gauge symmetries and K3 surfaces,''
Phys.\ Lett.\ B {\bf 357} (1995) 329
[hep-th/9507012].

\bibitem{DiDoGo}
D.~Diaconescu, M.~R.~Douglas and J.~Gomis,
``Fractional branes and wrapped branes,''
JHEP{\bf 9802} (1998) 013
[hep-th/9712230].

\bibitem{DiGo}
D.~Diaconescu and J.~Gomis,
``Fractional branes and boundary states in orbifold theories,''
JHEP{\bf 0010} (2000) 001
[hep-th/9906242].

\bibitem{Ta}
T.~Takayanagi,
``String creation and monodromy from fractional D-branes on ALE spaces,''
JHEP{\bf 0002} (2000) 040
[hep-th/9912157].

\bibitem{HuIeNuSc}
F.~Hussain, R.~Iengo, C.~Nunez and C.~A.~Scrucca,
``Interaction of moving D-branes on orbifolds,''
Phys.\ Lett.\ B {\bf 409} (1997) 101
[hep-th/9706186].

\bibitem{Sen98n}
A.~Sen,
``Stable non-BPS bound states of BPS D-branes,''
JHEP{\bf 9808} (1998) 010
[hep-th/9805019];

O.~Bergman and M.~R.~Gaberdiel,
``Non-BPS states in heterotic-type IIA duality,''
JHEP{\bf 9903} (1999) 013
[hep-th/9901014];

M.~R.~Gaberdiel and B.~J.~Stefanski,
``Dirichlet branes on orbifolds,''
Nucl.\ Phys.\ B {\bf 578} (2000) 58
[hep-th/9910109].

\bibitem{BiCrRo}
M.~Billo, B.~Craps and F.~Roose,
``On D-branes in type 0 string theory,''
Phys.\ Lett.\ B {\bf 457} (1999) 61
[hep-th/9902196].

\bibitem{Br}
I.~Brunner, R.~Entin and C.~Romelsberger,
``D-branes on T(4)/Z(2) and T-duality,''
JHEP{\bf 9906} (1999) 016
[hep-th/9905078].

\bibitem{BiCrRo2}
M.~Billo, B.~Craps and F.~Roose,
``Orbifold boundary states from Cardy's condition,''
JHEP{\bf 0101} (2001) 038
[hep-th/0011060].

\bibitem{BeDi}
M.~Bertolini, P.~Di Vecchia, M.~Frau, A.~Lerda, R.~Marotta and I.~Pesando,
``Fractional D-branes and their gauge duals,''
JHEP{\bf 0102} (2001) 014
[hep-th/0011077];

M.~Frau, A.~Liccardo and R.~Musto,
``The geometry of fractional branes,''
hep-th/0012035;

P.~Merlatti and G.~Sabella,
``World volume action for fractional branes,''
hep-th/0012193.

\bibitem{Ca}
J.~L.~Cardy,
``Boundary Conditions, Fusion Rules And The Verlinde Formula,''
Nucl.\ Phys.\ B {\bf 324} (1989) 581.

\bibitem{DiFrPeScLeRu}
P.~Di Vecchia, M.~Frau, I.~Pesando, S.~Sciuto, A.~Lerda and R.~Russo,
``Classical p-branes from boundary state,''
Nucl.\ Phys.\ B {\bf 507} (1997) 259
[hep-th/9707068].

\bibitem{deA}
S.~P.~de Alwis,
``Boundary string field theory the boundary state formalism and 
D-brane  tension,''
hep-th/0101200.

\bibitem{GiJo}
E.~G.~Gimon and C.~V.~Johnson,
``K3 Orientifolds,''
Nucl.\ Phys.\ B {\bf 477} (1996) 715
[hep-th/9604129].

\bibitem{BeGa2}
O.~Bergman and M.~R.~Gaberdiel,
``A non-supersymmetric open-string theory and S-duality,''
Nucl.\ Phys.\ B {\bf 499} (1997) 183
[hep-th/9701137].


\end{thebibliography}
\end{document}